%% file: andromeda18.tex
\documentclass[twocolumn,tighten]{aastex631}
\usepackage[starfontserif]{starfont}
\pdfoutput=1
\usepackage{graphicx,amsmath, natbib} 
\usepackage{url}
\usepackage[rightcaption]{sidecap}

\DeclareSymbolFont{starfontsym}{OT1}{sts}{m}{n}
\DeclareMathSymbol{\mathSun}{\mathord}{starfontsym}{115}

\newcommand{\andxviii}{And~XVIII\@}
\newcommand{\membersfeh}{38}
\newcommand{\membersvelo}{56}
\newcommand{\vhelio}{$-337.2$}

\newcommand{\feh}{{\rm [Fe/H]}}
\newcommand{\alphafe}{{\rm [$\alpha$/Fe]}}
\newcommand{\vheliotex}{v_{\rm helio}}
\newcommand{\vrot}{v_{\rm rot}}
\newcommand{\peff}{p_{\rm eff}}
\newcommand{\citeposs}[1]{\citeauthor{#1}'s (\citeyear{#1})}

\begin{document}

\title{Kinematics and metallicity of the dwarf spheroidal galaxy Andromeda XVIII}
\author[0000-0002-7438-1059]{Kateryna A. Kvasova}
\affiliation{Department of Physics and Astronomy, University of Notre Dame, 225 Nieuwland Science Hall, Notre Dame, IN 46556, USA}
\author[0000-0001-6196-5162]{Evan N. Kirby}
\affiliation{Department of Physics and Astronomy, University of Notre Dame, 225 Nieuwland Science Hall, Notre Dame, IN 46556, USA}
\author[0000-0002-1691-8217]{Rachael L. Beaton}
\affiliation{Space Telescope Science Institute, 3700 San Martin Drive, Baltimore, MD 21218, USA}

\begin{abstract}
Andromeda~XVIII is an isolated dwarf galaxy 579~kpc away from the nearest large galaxy, M31.  It is a candidate ``backsplash galaxy'' that might have been affected by a close passage to M31.  We present new Keck/DEIMOS spectroscopy of Andromeda~XVIII to assess the likelihood that it is a backsplash galaxy. We estimated the velocities, metallicities ([Fe/H]), and {\rm $\alpha$}-enhancements ({\rm [$\alpha$/Fe]}) for \membersvelo\ probable members. Based on the abundances of \membersfeh\ stars with low errors ($\delta {\rm [Fe/H]} < 0.3$), parameters for the simplest chemical evolution models were estimated using the maximum likelihood coupled with a Markov Chain Monte Carlo (MCMC) method.  The metallicity distribution is inconsistent with these models, due to a sharp metal-rich cutoff. We estimated Andromeda~XVIII's mean heliocentric velocity, rotation velocity, position angle of the rotation axis, and velocity dispersion using the maximum likelihood coupled with an CMC.  There is no evidence for bulk rotation, though subpopulations might be rotating.  The mean heliocentric velocity is \vhelio~km~s$^{-1}$, such that the line-of-sight velocity relative to M31 is lower than the escape velocity from M31. Together, the metallicity distribution and the mean velocity are consistent with a sudden interruption of star formation. For possible causes of this quenching, we considered gas loss due to ram pressure stripping during a close passage by M31 or due to a past major merger. However, we cannot rule out internal feedback (i.e., a terminal wind).

\end{abstract}


\section{Introduction}
\label{sec:intro}

The Milky Way (MW) and Andromeda (M31) are the two dominant galaxies in the Local Group. However, they are dominant only in luminosity and mass. By number, there are far more dwarf galaxies in the Local Group. Some of them are satellites of the MW or M31, and others are isolated. The MW has over 60 known satellites, and M31 has over 30. The Local Group hosts about 15 more isolated dwarf galaxies.\footnote{These numbers are subject to heavy observational bias, where it is easier to detect satellites of the MW compared to more distant dwarf galaxies.  See \citet{drl20} for an estimate of the luminosity function of MW satellites, corrected for selection bias.}

Any differences between the MW and M31 satellite populations seem insignificant when satellite dwarf galaxies are compared to isolated dwarf galaxies. Most obviously, the vast majority of satellite galaxies have no gas, and the vast majority of isolated galaxies have lots of gas \citep{spe14}, with typical gas-to-stellar mass ratios around 1. Isolated galaxies are also more likely to obey the radius--luminosity relation \citep{hig21} and the metallicity--luminosity relation \citep{kir13}. Altogether, the evidence suggests that dwarf galaxies experience a dramatic transition when they enter the sphere of influence of a large host, like the MW or M31. Their morphologies transition from dwarf irregular (dIrr) to dwarf spheroidal (dSph), and they lose their gas from ram pressure stripping. Galaxies with close enough pericentric passages can even lose dark matter and stellar mass to tidal stripping.

One intriguing class of isolated dwarf galaxies is the isolated dSphs. There are three famous isolated dSphs: Cetus, Tucana, and Andromeda XVIII (\andxviii). The prevalence of gas-rich isolated dwarf galaxies has led to the concept of ``backsplash galaxies'' \citep{tey12}. The orbits of backsplash galaxies once passed within the virial radius of a large galaxy, like the MW, even if those orbits are unbound. The brief interaction could be enough for ram pressure stripping to remove gas, quench star formation, and morphologically transform the galaxy from dIrr to dSph. The backsplash mechanism could be the proximate cause of the formation of supposedly dark matter-free, ultra-diffuse dwarf galaxies \citep{dan19,ben21}. Therefore, it is important to study the backsplash mechanism and its exemplars.

\begin{figure}
\centering
\includegraphics[width=1.\linewidth]{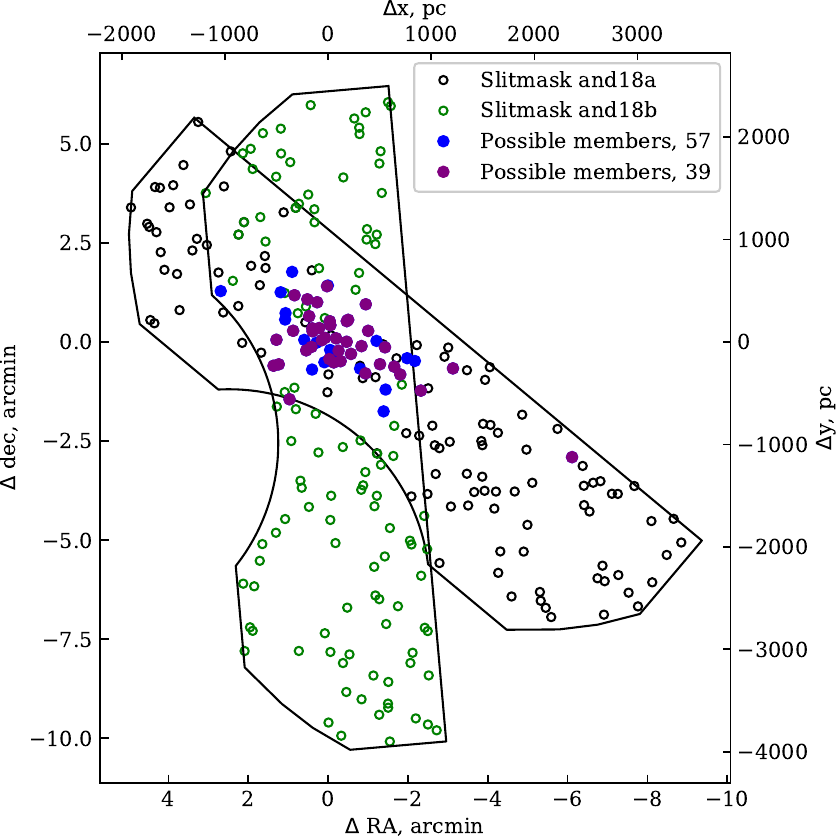}
\caption{Positions of objects observed via DEIMOS slitmasks and18a and and18b with surrounding objects. The probable members of \andxviii\ are shown in {\it blue} and {\it purple} (57 and 39 stars, respectively, including a metal-rich star PAndAS9654; see Section \ref{sec:members}).}
\label{fig:obs}
\end{figure}

\andxviii\ is a particularly interesting subject for further study.  \citet{mcc08} discovered and characterized \andxviii\ from CFHT/MegaPrime imaging.  However, a large portion of the galaxy was hidden by a CCD chip gap.  They found that the tip of the red giant branch (TRGB) has a magnitude $I_0=21.6$.  \citet{mak17} later found a TRGB magnitude of ${\rm F814W}=21.7$ from HST/ACS photometry.  \citet{tol12} obtained the first spectra of stars in \andxviii\@.

It is not likely to be a backsplash galaxy of the MW, because its Galactocentric distance is about $1.33$ Mpc. On the other hand, its 3-dimensional distance from M31 is only $579$ kpc \citep{mak17}. Therefore, it could be an example of an M31 backsplash dwarf galaxy. Its photometrically measured star formation history (SFH) shows two epochs of intense star formation: one $12-14$ Gyr ago, and another that started about $8$ Gyr ago, with a sudden cessation about $1.5$ Gyr ago \citep{mak17}. The second star formation period might be associated with a flyby of M31.

The purpose of this work is to determine the features of the evolution of \andxviii\ using accurate measurements of kinematical properties, metallicities and $\alpha$-enhancement of the most probable \andxviii\ members, together with chemical evolution models. Specifically, we determine here the role of gas loss -- ram pressure stripping -- based on chemical properties, but also the presence or absence of rotation and the line-of-sight dwarf galaxy's velocity relative to M31, to find whether it can exclude the possibility of \andxviii\ being a backsplash dwarf.  


\section{Observations}
\label{sec:observations}

We observed \andxviii\ with the Deep Imaging Multi-Object Spectrograph \citep[DEIMOS,][]{fab03} on the Keck II telescope on 2022 September 19--22. We designed two slitmasks: and18a and and18b.  The total exposure times were 590 and 556 minutes, respectively, for and18a and and18b.  The conditions were excellent, with seeing ranging between {\rm $0.''5$} and {\rm $0''.8$}.  The paragraphs below describe the target selection.
 
Each slit mask has an approximate sky projection of $16\arcmin \times 5\arcmin$. We used the OG550 order-blocking filter with the 1200G grating at a central wavelength of 7800~\r{A}\@.  The slit widths were 0.8$\arcsec$, resulting in a spectral resolving power of $R \sim 6000$, or a line width of 1.3~\r{A}\ FWHM\@. We used the \texttt{spec2d} pipeline from the DEEP2 Galaxy Redshift Survey \citep{coo12,new13} to reduce the raw frames.  The result of the pipeline is a list of one-dimensional, sky-subtracted, wavelength-calibrated spectra.  We used some improvements to the wavelength calibration and one-dimensional object extraction described by \citet{kir15c,kir15b}.

We selected targets from three different catalogs.  First, we selected stars from photometry acquired with the \textit{Hubble Space Telescope} Advanced Camera for Surveys \citep[SNAP project 13442, PI: R.~B.\ Tully;][]{mak17}.  And~XVIII was observed with the F606W and F814W filters.  L.\ Makarova kindly shared with us the photometric catalog, which was the output of DOLPHOT \citep{dol16}.  The catalog had pixel positions, which we converted to sky coordinates using the world coordinate solution of the FITS files of the observations available from the MAST archive.

The ACS field of view is significantly smaller than a DEIMOS slitmask.  Therefore, we also included photometry from the Pan-Andromeda Archaeological Survey \citep[PAndAS,][]{iba14,mcc18}.  PAndAS is a wide-field survey of M31 and M33 with CFHT/MegaCam in the $g$ and $i$ filters.  The survey has gaps between CCDs.  Unfortunately, one of these gaps coincided with the center of And~XVIII\@.  Therefore, PAndAS photometry is useful only for stars in the outer regions of the dwarf galaxy.

\begin{figure*}
\centering
\includegraphics[width=1.\linewidth]{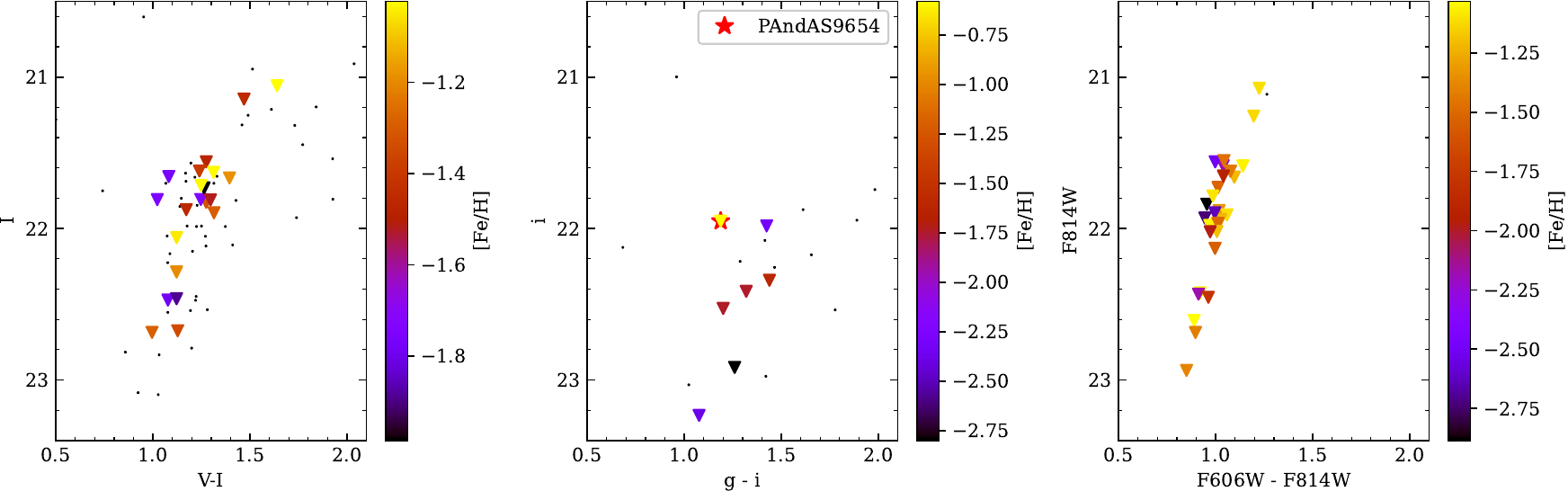}
\caption{Color-magnitude diagrams (CMD) for 56 probable \andxviii\ members plus PAndAS9654, assumed to be an M31 halo star (see Section \ref{sec:members}), in three photometric systems: KPNO ({\it left}), PAndAS ({\it middle}), and HST/ACS ({\it right}). The color gradient shows the {\rm [Fe/H]} values for the stars. Although PAndAS9654 is on the left side of the CMD in the middle panel, its spectroscopic metallicity is the largest in our sample. According to Figure~\ref{fig:gradient_dist}, this star is also {\rm $\alpha$}-enhanced compared to the other 56 objects.}
\label{fig:cmd3}
\end{figure*}

Photometry from the Spectroscopic and Photometric Landscape of the Andromeda Stellar Halo (SPLASH) project was also used. 
The imaging was obtained on the MOSAIC 1.1 instrument on the Kitt Peak National Observatory (KPNO) Mayall 4-meter telescope using the Washington filter system \citep[$M$, $T_2$;][]{1976AJ.....81..228C} supplemented with the $DDO51$ filter that is sensitive to stellar surface gravity for late-type stars \citep{maj00} (NOIRLab Prop. ID  2010B-0596; PI: R.~Beaton). 
MOSAIC had a field-of-view of $36' \times 36'$ using eight 2k~$\times$~4k CCDs with chip gaps. And~XVIII was placed on a single CCD for uninterrupted coverage of the dwarf galaxy. 
Image processing used the \texttt{mscred} package in IRAF \citep[a detailed description is given in][]{2014PhDT.......416B}. 
The photometry was obtained using the DAOPHOT family of programs \citep{1987PASP...99..191S,1990PASP..102..932S,1994PASP..106..250S} following the procedures used for the broader SPLASH survey \citep{2014PhDT.......416B,2003PhDT.........4O} and used other SPLASH spectroscopic follow-up \citep[e.g.,][]{2007ApJ...670L...9M,tol12}.
The photometry was calibrated directly to Landolt standards \citep{1996AJ....111..480G, lan92} or, for data taken in non-photometric conditions, bootstrapped to observations on photometric nights.
Image astrometry used USNO-B \citep{2003AJ....125..984M}. 
We converted $M$ and $T_2$ magnitudes to the Johnson--Cousins system based on the transformations of \citet[][Section~2.2]{maj00}. 

We supplemented the above catalogs with photometry from the \textit{Gaia} DR3 catalog.  The \textit{Gaia} catalog contains very few And~XVIII members because they are too faint.  Nonetheless, the catalog is still useful for slitmask alignment stars.

We matched stars from the different catalogs based on their coordinates.  Then, we assigned priorities for placement on the slitmasks to the stars based on their position in the color--magnitude diagram (CMD), and in some cases, from the color--color diagram involving the Washington $DDO51$ filter.  We drew a generous selection polygon in the three CMDs defined by the HST/ACS, PAndAS/CFHT, and KPNO (transformed to Johnson--Cousins) filter sets.  Stars outside of the polygons were considered nonmembers.  Within the selection boxes, the stars were prioritized by magnitude.  First, stars with brighter HST/ACS F814W magnitudes were given higher priority.  Then, for stars not observed with HST, stars with brighter CFHT magnitudes were given higher priority.  Finally, for stars observed with neither HST nor CFHT, stars with brighter KPNO magnitudes were given higher priority.

The $DDO51$ filter brackets the Mg~b triplet, which is sensitive to surface gravity \citep{maj00}.  Dwarf stars have strong Mg~b triplet features and are therefore fainter in the $DDO51$ filter relative to giant stars of the same $M-T2$ color (as a proxy for temperature).  Washington color--color diagrams, specifically $M-T2$ versus $M-DDO51$ \citep[see e.g.,][]{maj00,tol12}, can separate foreground dwarf stars in the Milky Way from target giants in the M31 system (including And~XVIII)\@. 
The strength of the Mg~b triplet varies with stellar temperature, and dwarf stars in the Milky Way fall on a characteristic ``swoosh'' shape in the color--color diagram \citep[see][their Figure~6]{maj00}. The giant probability, or $\rm{gprob}$, \citep{2003PhDT.........4O} is the fraction of a source's color--color error circle within the region populated by giants \citep[for an example, see Figure 2 of][]{tol12}.
Following other SPLASH methodologies, we considered stars with $\rm{gprob} < 0.1$ as nonmembers.  They were given priorities as low as stars that fell outside of the CMD selection polygons.

We created the slitmask designs with the \texttt{dsimulator} software.  Figure~\ref{fig:obs} shows the slitmasks projected onto the sky.  Figure~\ref{fig:cmd3} shows the stars' color--magnitude diagrams in the filter sets of the various source catalogs.


\section{Methods}
\label{sec:methods}

We measured radial velocities (Section~\ref{sec:rv}) and abundances (Section~\ref{sec:abund}) from the DEIMOS spectra.  In order to make these measurements, the spectra needed to be prepared for analysis (Section~\ref{sec:spec}).

\subsection{Preparation of Spectra}
\label{sec:spec}

We divided each spectrum by a telluric standard spectrum following the procedure of \citet[][Section~3.2]{kir08}.  The standard star was HR7346, a rapidly rotating star of spectral type B9V\@.  The star was observed on 2015 May 19 by drifting the star across a {\rm $0.''7$} slit during the exposure.
The spectrum was continuum-normalized.  The absorption depths were scaled to the airmass of the And~XVIII observations before division. 

\input{and18_10_objects_main.tab}

Still, some wavelength windows are severely disturbed by telluric absorption, and we excluded them from the measurements of radial velocity and elemental abundances \citep{kir08}. However, these regions are still included in the slit miscentering correction described in Section~\ref{sec:rv}.

\subsection{Radial Velocities}
\label{sec:rv}

We measured radial velocities by cross-correlating each spectrum with a suite of template spectra.  \citet{kir15c} described the observations of the template spectra, as well as the procedure for performing the cross correlation, which was based on the work of \citet{sim07}.  The uncertainties in radial velocity were estimated from Monte Carlo resampling of the spectra.  The redshift corresponding to this radial velocity, $z_{\rm rest}$, is used to place the spectra in the rest frame.

The velocities include a correction for slit miscentering, which can shift the zeropoint of a spectrum's wavelength by up to $\sim 20$~km~s$^{-1}$.  This correction is not necessary to shift the spectra into the rest frame, but it is necessary to measure absolute velocities with respect to a standard of rest.  We followed the procedure of \citet{soh07}, which is further expanded upon by \citet{kir15c}.  In summary, the telluric A and B bands are cross-correlated with the same spectral region in the telluric standard star.  To compute the stellar velocities in the geocentric frame, the inferred ``redshift'' of the telluric features is added to the observed redshift of the star.

We shifted the velocities to the heliocentric reference frame based on the star's coordinates and mean observation time.  The heliocentric velocity for a star is called $v_{\rm helio}$.  We use these velocities in a procedure to infer kinematic properties of \andxviii\ in Section~\ref{sec:kin}, whose results are presented in Section \ref{sec:resvel}.  

\subsection{Elemental Abundances}
\label{sec:abund}

Our approach is based on that of \citet{kir08,kir09,kir10} and \citet{esc19}. We derived the effective temperature $T_{\rm eff}$, surface gravity $\log g$, metallicity [Fe/H], and {\rm $\alpha$}-enhancement [$\alpha$/Fe] via spectral synthesis. We measured an averaged $\alpha$ abundance, but we did not measure individual $\alpha$ element abundances. An alternative approach with low- to medium-resolution spectroscopy is an empirical relation between abundance and line strengths, such as the \ion{Ca}{2} triplet.  However, those relations possibly confuse the abundances of different elements.  For example, \ion{Ca}{2} triplet strength could depend on both the Ca and Fe abundances, where Fe abundance is a proxy for free electron fraction and therefore the overall metallicity \citep{bat08,sta10}.  Strong lines like the \ion{Ca}{2} triplet are poorly modeled in local thermodynamic equilibrium (LTE)\@.  Our spectral synthesis approach assumes LTE, so we exclude such lines.  \citet{kir08, kir09, kir10} list the rest-frame spectral regions considered within our analysis.

In summary, we measured the aforementioned four stellar parameters by matching the observed spectra to a grid of synthetic spectra generated using the LTE spectral synthesis code MOOG \citep{sne73, sne12}. This approach is especially useful for spectra of distant stars -- such as those in \andxviii\ -- whose spectra have low S/N\@.  Although individual lines may be only poorly detected, the ensemble of many lines has enough signal to permit measurements of [Fe/H] and [$\alpha$/Fe].  The spectra were synthesized in model atmospheres generated with ATLAS9 \citep{kur17}.  The grid of atmospheres and spectra models is described by \citet{kir11d}.

In order to make optimal use of all available data, we used broadband photometric magnitudes and colors to estimate $T_{\rm eff}$ and $\log g$.  We interpolated a model isochrone using the known color and magnitude of each star, combined with the distance modulus of And~XVIII \citep[$(m-M)_0 = 25.62$;][]{mak17}. PARSEC isochrones\footnote{ The isochrones for all three photometric systems were generated using the software CMD 3.7: \url{http://stev.oapd.inaf.it/cgi-bin/cmd}} \citep{gir02} were used for all three photometric systems of observed objects: Johnson-Cousins (KPNO), CFHT (PAndAS), and HST WFC (ACS). We assumed an age of 12~Gyr.  The values of $T_{\rm eff}$ and $\log g$ are not very sensitive to the choice of isochrone age.

With $T_{\rm eff}$ and $\log g$ fixed to their photometric values, we used $\chi^2$ minimization to fit each observed spectrum to the interpolated grid of synthetic spectra. For continuum normalization during the fit, we used third-order B spline interpolation.  We began with an initial continuum that was iteratively refined for specific regions of spectra \citep{kir08}. The breakpoint spacing was chosen to be 100 pixels in order to avoid overfitting. We used {\rm 5$\sigma$} and {\rm 3$\sigma$} clipping for the initial and refined continua, respectively. The best-fitting values of \feh\ and \alphafe\ comprise our measurements of those parameters for each star.

[$\alpha$/Fe] abundances were calculated from a simultaneous fit to Mg, Si, Ca, and Ti wavelength masks, following the procedure of \citet{kir08}. We do not report abundances of individual $\alpha$ elements. Generally low S/N (e.g., more than half of estimated members have ${\rm S/N}<15$) would result in large errors for these individual abundances.

Table~\ref{tab:mrt} gives object names, slitmasks, S/N, heliocentric velocities, coordinates (ICRS), membership, and also {\rm T$_{\rm eff}$}, surface gravity, \feh, \alphafe, and random errors for some of the stars that we observed. The table contains only stars with {\rm $v_{\rm helio}<100~{\rm km~s}^{-1}$} and {\rm $v_{\rm helio}>-400~{\rm km~s}^{-1}$} (in total 166 objects). 

\begin{figure*}
\centering
\includegraphics[width=0.48\linewidth]{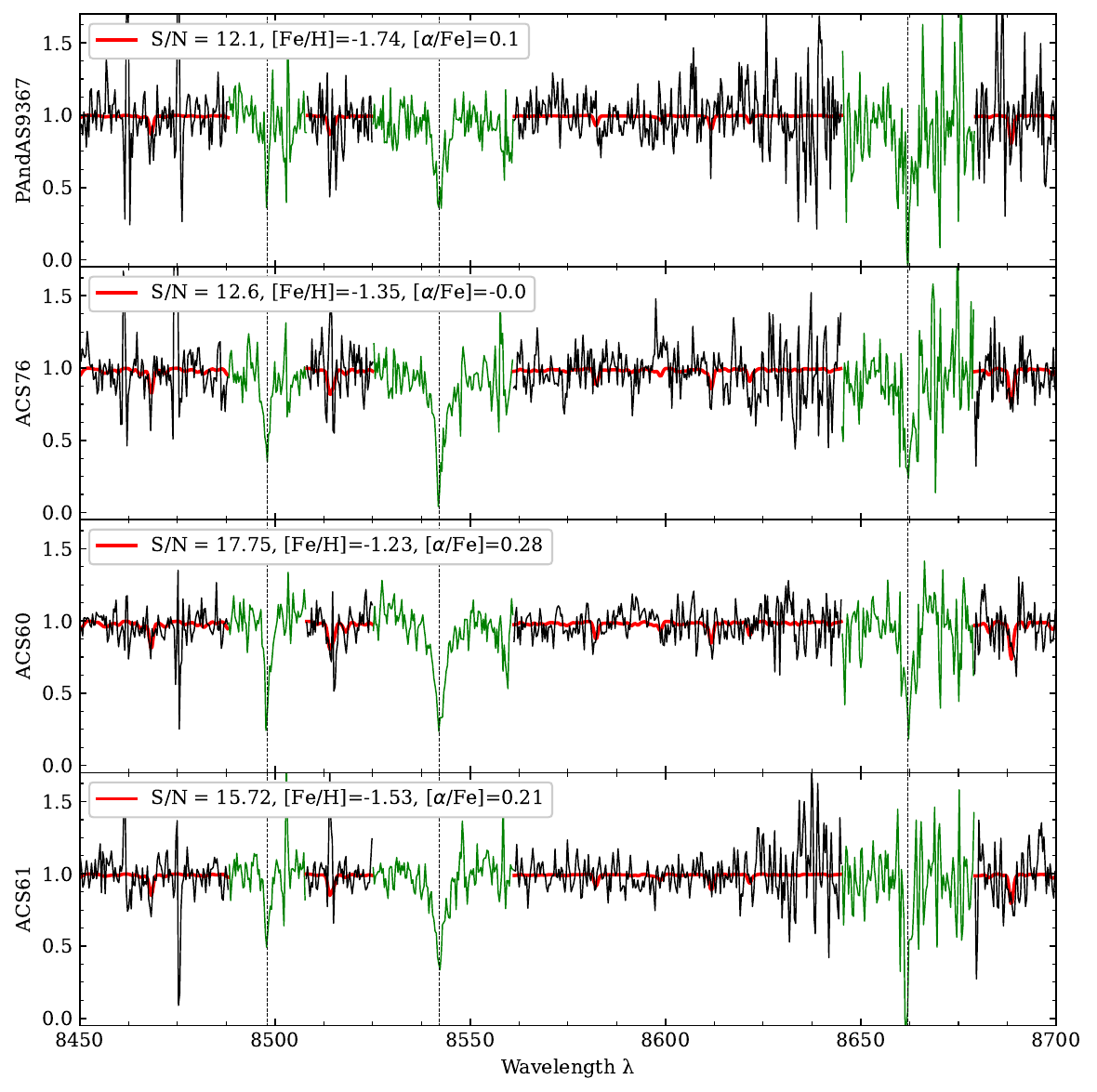}
\includegraphics[width=0.48\linewidth]{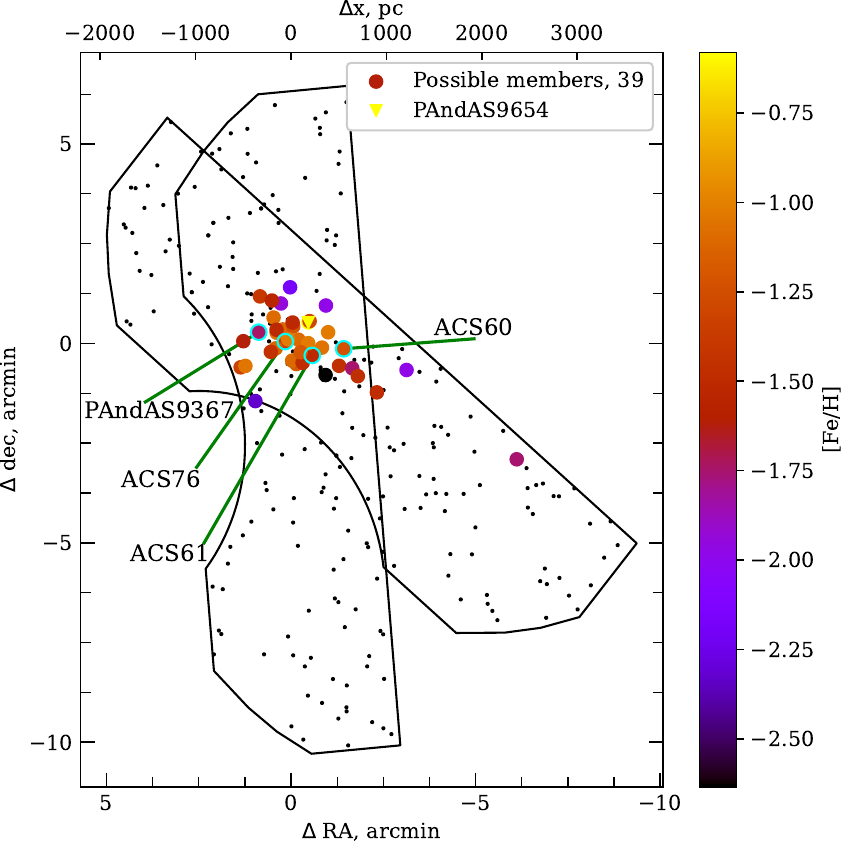}
\caption{{\it Left:} Examples of member stars spectra with different S/N. Figure~\ref{fig:gradient_ra} (right) contains their subscripts. {\it Black} is used to indicate normalized flux; {\it green} indicates spectral regions excluded from abundance analysis; {\it red} is for interpolated synthetic spectra with \feh\ and \alphafe\ as indicated in the plot's legend; and vertical lines mark the Ca triplet lines. {\it Right:} Map of \andxviii\ stars color-coded by metallicity.  Small black points are nonmembers. The two DEIMOS slitmasks are shown with outlines.}
\label{fig:gradient_ra}
\end{figure*}

\noindent
Nonmembers are at different observed velocities than \andxviii, while the isochrone method for {\rm T$_{\rm eff}$} and surface gravity was used with the same distance modulus, which caused us to classify PAndAS9654 as an M31 halo star, not a member of \andxviii\ (Figure~\ref{fig:cmd3}).

Although most stars have measured abundances, only those with $\delta{\rm [Fe/H]}<1$ are included in Table~\ref{tab:mrt}. For membership criteria, see Section~\ref{sec:members}.

\input{2_main.tab}

\subsection{Analytic Chemical Evolution Models}
\label{sec:CEM}

The measured metallicity distribution (N$_{\rm stars} $ in bins of [Fe/H]) of a galaxy can be described by chemical evolution models. These models are used for a general assessment of the details of star formation, interaction with the environment, and the ratio of the mass of gas to the mass of stars. The Leaky Box, Pre-Enriched, Accretion, and Ram Pressure Stripping models (\citealt{kir13} and references therein) were used in this work (see Table~\ref{tab:results}). All of them are one-zone models, and they share simple assumptions, including instantaneous recycling of material and the Kennicutt--Schmidt law for the star formation rate.

\subsubsection{Leaky Box Model}
\label{sec:lb}
The Leaky Box model assumes that all stars in the galaxy form from gas with a primordial composition.  The galaxy is subject to gas loss (``leaking'') due to internal feedback. This also can be called the Pristine model. Its metallicity distribution is described by the formula
\begin{equation}
\frac{dN}{d{\rm [Fe/H]}}\propto\left(\frac{10^{\rm [Fe/H]}}{\peff}\right)\exp\left({\frac{-10^{\rm [Fe/H]}}{\peff}}\right)
\label{eq:1}
\end{equation}
where $\peff$ is the effective yield.  The effective yield is $\peff=p/(1+\eta)$, where the true yield $p=Z/(\ln(\mu^{-1}))$ is in units of the solar metal fraction ($Z_\mathSun$). $\eta$ is the mass loading factor, or the proportionality coefficient between outflow and star formation rate (SFR); $Z$ is the current gas-phase metallicity; and $\mu$ is the current gas fraction.

\subsubsection{Pre-Enriched Model}
\label{sec:enri}
The Pre-Enriched model assumes an enriched composition of the initial gas. Its metallicity distribution is described by:
\begin{equation}\begin{split}
\frac{dN}{d{\rm [Fe/H]}}\propto\left(\frac{10^{\rm [Fe/H]}-10^{{\rm [Fe/H]}_{0}}}\peff\right)
\times\exp\left({\frac{-10^{\rm [Fe/H]}}{\peff}}\right)
\label{eq:2}
\end{split}\end{equation}

${\rm [Fe/H]}_0$ describes the initial galaxy's gas metallicity. For very low values (${\rm [Fe/H]}_0\lesssim-5$) the Pre-Enriched model reduces to the Leaky Box model.

\subsubsection{Accretion Model}
\label{sec:accr}
The Accretion model contains three parameters: the effective yield $\peff$, the ratio of the final mass to the initial (gas) mass $M$, and the stellar mass fraction $s$, which is determined from the numerical solution of the following equation:
\begin{equation}\begin{split}
{\rm [Fe/H]}(s) = \log\Biggl[\peff\left(\frac{M}{1+s-s/M}\right)^2\\
\times\left(\ln\left(\frac{1}{1-s/M}\right)-\frac{s}{M}\left(1-\frac{1}{M}\right)\right)\Biggr]
\label{eq:3}
\end{split}\end{equation}
\begin{equation}\begin{split}
\frac{dN}{d{\rm [Fe/H]}}\propto\left(\frac{10^{\rm [Fe/H]}}\peff\right)\Biggl[1+s\left(1-\frac{1}{M}\right)\Biggr]\\
\times{\Biggl[\left(1-\frac{s}{M}\right)^{-1}-2\left(1-\frac{1}{M}\right)\left(\frac{10^{\rm [Fe/H]}}{\peff}\right)\Biggr]}^{-1}
\label{eq:4}
\end{split}\end{equation}

\subsubsection{Ram Pressure Stripping Model}\label{sec:rps}
The Ram Pressure Stripping model describes the evolution of the dwarf galaxy that lost its gas due to a close passage by the host galaxy.  The motion of such a  galaxy through the comparatively dense CGM of the host galaxy exerts a pressure on the dwarf galaxy that expels its gas. The model depends on three parameters, namely $\peff$, the effective yield; \feh$_s$, the metallicity of the galaxy at the moment that gas loss commences; and $\zeta$, a parameter that describes the intensity, expressed as a fraction of the ram pressure gas loss in comparison to the gas loss caused by supernova winds (see~\citealt{kir13}). 
\begin{equation}\begin{split}
\frac{dN}{d{\rm [Fe/H]}}\propto\left(\frac{10^{\rm [Fe/H]}}{\peff}\right) \Biggl[\exp\left({\frac{-10^{\rm [Fe/H]}}{\peff}}\right)\\
+ \zeta \left(\exp\left({\frac{10^{{\rm [Fe/H]}_{s}}-10^{\rm [Fe/H]}}{\peff}}\right)-1\right)\Biggr]
\label{eq:50}
\end{split}\end{equation}

\input{3_mcmc.tab}

\subsubsection{Likelihood for Chemical Evolution Models}

To find the most likely parameters of each model, the likelihood was calculated according to the formula
\begin{equation}\begin{split}
L=\prod_i \int_{-\infty}^{\infty} \frac{dP}{d{\rm [Fe/H]}} \frac{1}{\sqrt{2\pi}\cdot\delta{\rm [Fe/H]}_i}\\
\times\exp{\left(-\frac{\left({\rm [Fe/H]}-{\rm [Fe/H]}_i\right)^2}{2\left(\delta{{\rm [Fe/H]}}_i\right)^2}\right)}\,d{\rm [Fe/H]}
\label{eq:5}
\end{split}\end{equation}

For convenience, the logarithm of $L$ was maximized. The maximization was implemented using the Markov Chain Monte Carlo method of the $emcee$ python library (Section~\ref{sec:mcmc}).

Each metallicity error includes systematic error as $\delta {\rm [Fe/H]}_i=\sqrt{\delta {\rm [Fe/H]}_{i,{\rm rand}}^2+0.106^2}$ \citep{kir10}, where $\delta{\rm [Fe/H]}_{i,{\rm rand}}$ is the random uncertainty from the spectral fit \citep{kir15c}.

\subsection{Kinematics}
\label{sec:kin}

We assumed a Gaussian distribution of velocities for \andxviii\@.  Although binary stars can inflate the measured velocity dispersion, we ignored them because we do not have multi-epoch spectroscopy. The mean heliocentric velocity and velocity dispersion were estimated using a Markov Chain Monte Carlo (MCMC) method to maximize the likelihood of the galaxy that is described by a normal distribution:
\begin{equation}\begin{split}
\ln L = -\frac{1}{2} \sum_{i}^{N} \ln \left( 2\pi \left({(\delta v)_i}^2 +{\sigma_v}^2\right)\right)\\
-\frac{1}{2}\sum_i^{N}\Biggl[\frac{{\left((\vheliotex)_i-\langle \vheliotex\rangle\right)}^2}{{(\delta v)_i}^2+{\sigma_v}^2}\Biggr]
\label{eq:6}
\end{split}\end{equation}

We also considered rotation about an axis $\theta$ with a separate likelihood function:
\begin{equation}\begin{split}
\ln L = -\frac{1}{2} \sum_{i}^{N} \ln \left( 2\pi \left({(\delta v)_i}^2 +{\sigma_v}^2\right)\right) \\
-\frac{1}{2}\sum_i^{N}\Biggl[\frac{{\left((\vheliotex)_i-\left(\langle \vheliotex\rangle+\vrot\cos{(\theta-\theta_i)}\right)\right)}^2}{{(\delta v)_i}^2+{\sigma_v}^2}\Biggr]
\label{eq:7}
\end{split}\end{equation}

The index of each star is $i$.  The angles are measured with respect to north.  Thus, $\theta_i$ is the angle between star $i$ and the center of the galaxy.  The free parameters are the mean velocity $\langle v_{\rm helio} \rangle$, the velocity dispersion $\sigma_v$, the rotation velocity $v_{\rm rot}$, and the position angle of the rotation axis $\theta$. Each star's radial velocity error includes systematic error as $\delta v_i=\sqrt{\delta v_{i,obs}^2+(1.49~{\rm km~s}^{-1})^2}$ \citep{kir15c}.

\subsection{MCMC Ensemble Sampler}
\label{sec:mcmc}

In this work we used the MCMC ensemble sampler method to find the most probable parameters of the likelihood functions constructed in Sections~\ref{sec:CEM} -- \ref{sec:kin}. 

We used the MCMC sampler $emcee$\footnote{\url{https://pypi.org/project/emcee/}}. The mathematical derivation and details of the method are from \citet{goo10} and \citet{for13}. 
Those authors recommend choosing the number of walkers to be larger than the number of free parameters in the model. The number of parameters, walkers, iterations, $\log(L)$, and corrected Akaike information criterion \citep[AICc;][see Section \ref{sec:reschem}]{aka74,sug78} are given in Table~\ref{tab:mcmcwalkers}.

With $emcee$, we measured the mean integrated autocorrelation time every 100 iterations. For the convergence criterion, we required the change of the autocorrelation time relative to the previous measurement to be less than 0.01.  The number of links in the chain is not less than 100 times the derived time. Further details on autocorrelation derivation for ensembles of chains are described by \citet{goo10} and \citet{for13}.


\section{Results}
\label{sec:res}

\subsection{Membership}
\label{sec:members}

In total more than 170 objects were observed. However, most of these are nonmembers (some by design) or have spectral quality insufficient for measurement of velocity and/or metallicity.

Only 100 objects have sufficient S/N and quality to measure abundances. Next, 30 objects were excluded due to the presence of strong sodium 8195~\r{A}\@ lines. The lines are used as an indicator of high surface gravity, such as found in foreground dwarf stars \citep[e.g.,][]{gil06}. Figures~\ref{fig:veldist} and \ref{fig:velhist} show that all of these stars have radial velocity values far from that of And~XVIII\@. After that, 13 objects were excluded based on the velocity criterion: their $v_{\rm helio}$ were more than 3$\sigma_v$ away from $\langle v_{\rm helio} \rangle$. Here we repeated the method used by \citet{kir10}. The method consists of iteratively estimating $v_{\rm helio}$ and $\sigma_{v}$.  Each iteration considers only objects within 3$\sigma_v$ around $\langle v_{\rm helio} \rangle$ (as determined in the previous iteration), until the number of accepted members does not change. For the purpose of fitting chemical evolution models, we additionally excluded stars that have either \feh\ random uncertainties or \alphafe\ random uncertainties that exceed 0.3.

There are 56 objects that are \andxviii\ members.  The 57th object is the most metal-rich star in our sample.  We argue below that it is a member of M31's halo, not that of \andxviii\@.  The membership list for the derivation of the chemical evolution models' parameters contains 38 stars after excluding the most metal-rich star.

\subsubsection{Contamination from M31 Halo Stars}

It is impossible to completely remove foreground and background objects with 100\% confidence. In particular, since PAndAS9654 has a higher metallicity than the other 38 members, we discuss our reasons for counting it as a member of the M31 halo, rather than that of \andxviii\@. 

This star is shown in Figure~\ref{fig:gradient_dist} as the most metal-rich star with {\rm \feh~=~$-0.575$}, $[\alpha/{\rm Fe}] = 0.503$, and in Figure~\ref{fig:gradient_ra} as a yellow star symbol. Figure~\ref{fig:cmd3} shows how PAndAS9654 is distinct from members of \andxviii\@.  Metal-rich stars are expected to be on the red side of the red giant branch (RGB), but PAndAS9654 appears on the blue (left) side of \andxviii's RGB\@.  This position would be expected if it is a less luminous star but at a lower distance modulus than \andxviii\@.

\begin{figure}
\centering
\includegraphics[width=1.\linewidth]{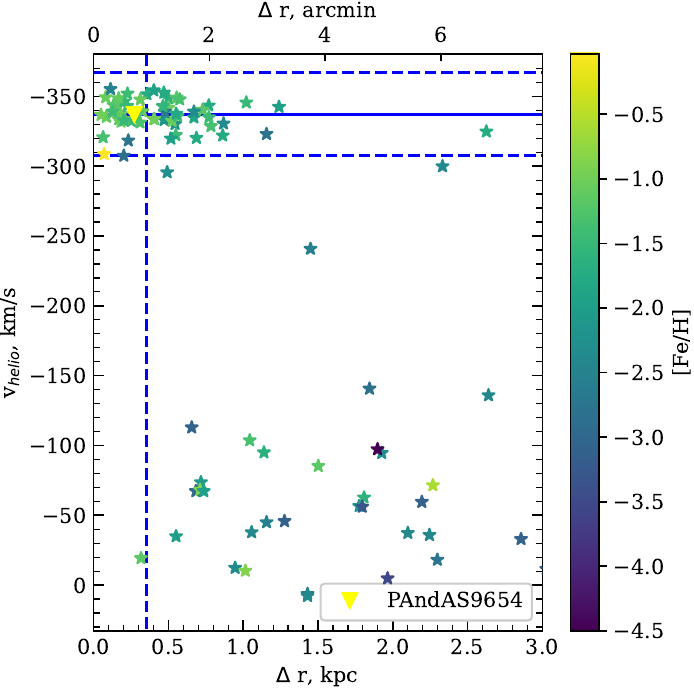}
\caption{Radial velocity vs.\ projected distance from the center of \andxviii\@. Horizontal lines correspond to  $\langle v_{\rm helio} \rangle$ of the dwarf galaxy and a {\rm 3$\sigma_v$} region around it. The vertical line corresponds to the half-light radius (\rm $r_{1/2}$) calculated with the \citeposs{mcc08} angular value and \citeposs{mak17} distance (see Table~\ref{tab:comparison}). The reader may compare the plot with Figure~12 of \citet{tol12} from the SPLASH overview of M31 satellite system.}
\label{fig:veldist}
\end{figure}

\begin{figure}
\centering
\includegraphics[width=1.\linewidth]{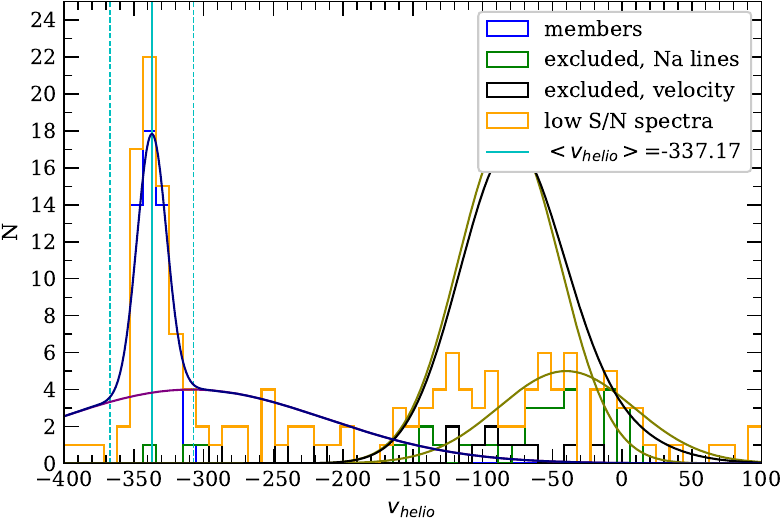}
\caption{Velocity distribution. The objects that were excluded due to the presence of strong \ion{Na}{1}~8195 lines are shown in {\it green}. The objects that were excluded due to the velocity criterion are shown in {\it black}. The {\it yellow} histogram includes the 170 objects from both and18a and and18b slitmasks within the radial velocity range shown. Also shown are the approximate shapes of velocity distributions for MW stars (combination of the disk and halo with  $\langle v_{\rm mw1} \rangle = -81.2$~km s$^{-1}$, $\sigma_{\rm mw1}$ = 36.5 km s$^{-1}$, $\langle v_{\rm mw2} \rangle= -40.2$~km s$^{-1}$, $\sigma_{\rm mw2}$ = 48.5 km s$^{-1}$) and the Andromeda halo ($\langle v_{\rm M31} \rangle = -308.8$~km s$^{-1}$, $\sigma_{\rm M31}$ = 96.3 km s$^{-1}$; \citealt{col13}) on the plot.  The approximate fraction of M31, MW, and \andxviii\ stars among 170 objects for Gaussians were derived using \citeposs{col13} approach. The normalization of the velocity distribution is arbitrary.}
\label{fig:velhist}
\end{figure}

\begin{figure*}
\centering
\includegraphics[width=0.48\linewidth]{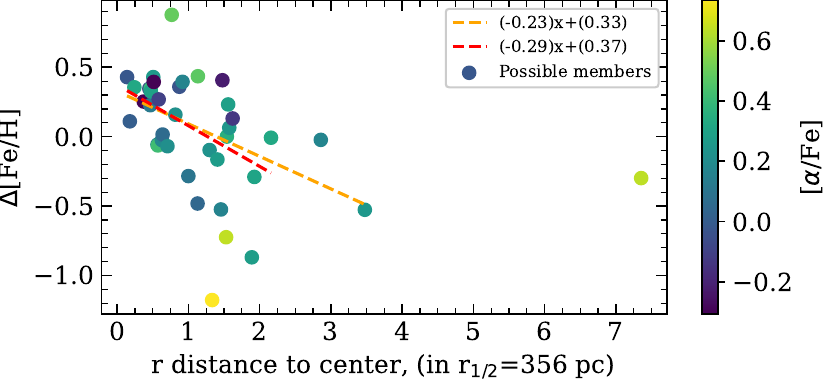}
\includegraphics[width=0.48\linewidth]{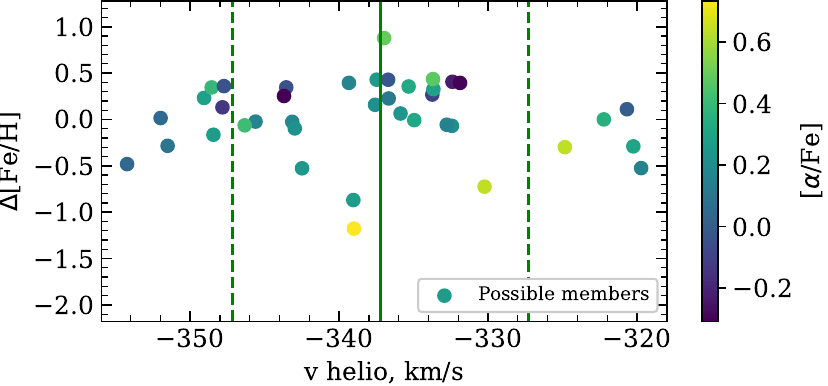}
\caption{{\it Left:} Metallicity distribution with the distance to the center color-coded by [$\alpha$/Fe]. The point with the largest \feh\ is the star PAndAS9654 (see Section~\ref{sec:members}). {\it Right:} Metallicity vs.\ $v_{\rm helio}$ color-coded by [$\alpha$/Fe]. Here, $v_{\rm helio}$ and $\sigma_v$ are shown with {\it solid} and {\it dashed} lines, respectively.}
\label{fig:gradient_dist}
\end{figure*}

Therefore, in order to further consider its membership in \andxviii, we tried the velocity method used by \citet{col13}. The authors developed the method to consider the probability that each star belongs to one of three velocity Gaussians: the MW, M31's halo, and the dwarf galaxy. This approach is reasonable, but it is limited by the strong overlap between the Gaussians for M31's halo and \andxviii\@. Figure~\ref{fig:veldist} shows how {\rm PAndAS9654} overlaps with \andxviii\ radial velocity and spatial distribution, while Figure~\ref{fig:velhist} shows the velocity distribution of 170 observed objects along with Gaussians that represent the mean and standard deviation of velocities in M31, the MW, and \andxviii\ (the rest of the observed objects have radial velocities outside of the plot's range: $v_{\rm helio}~<~-400$ km s$^{-1}$ or $v_{\rm helio}~>~100$~km~s$^{-1}$). All M31 and MW halo parameters that we used were taken from \citet{col13}. The normalization of the Gaussians in the figure is arbitrary. The metal-rich star already passed the velocity membership test. To sum up, the velocity criterion doesn't solve all of  the difficulties in determining secure dwarf galaxy's stars.

For this reason, \citet{tol12} used the set of membership criteria that does not include velocity and also emphasized that it is difficult to distinguish M31 halo stars from satellite members. The authors mentioned the importance of metallicity differences between satellite populations of M31 (${\rm [Fe/H]} \approx -1.4$ to $-2.0$) and M31 halo stars (${\rm [Fe/H]} \approx -1.5$ to $-0.1$), although they used photometrically derived metallicities. A more thorough way to consider membership in the M31 halo was studied by \citet[][and references therein]{gil12}. Based on their success rate of recovering M31 members as a function of distance from M31's center, a DEIMOS observer may expect about 1--3 M31 halo stars per slitmask at a projected radius of $\approx$114~kpc, which corresponds to \andxviii\ (see Appendix~\ref{sec:app2b}). \andxviii's three-dimensional distance is 579 kpc {\it behind} M31. Therefore, we must look through the entirety of M31's halo to see \andxviii, thus increasing the possibility of contamination by halo stars. 

These arguments support our assessment that PAndAS9654 is more likely a member of M31 than of \andxviii\@. It was the only star that we additionally excluded as an M31 halo member, because its CMD position is inconsistent with its spectroscopic metallicity, and its metallicity is much higher than any other member star (Figs.~\ref{fig:cmd3}, \ref{fig:gradient_ra}, and \ref{fig:gradient_dist}). For that reason, we estimated kinematical properties for both datasets -- with and without PAndAS9654 -- to compare its impact on the results.

At the same time, we are not excluding any other metal-rich stars from their CMD position. Specifically, \citet{boy15} and \citet{boy19} obtained infrared photometry of some Local Group dwarf galaxies, including \andxviii\@. They statistically derived upper limits for the numbers of TRGB and approximate giant branch (AGB) stars. \citeauthor{boy15}\ found that \andxviii\ appears to have one of the largest populations of AGB stars, although \citeauthor{boy19}\ later found no pulsating AGB stars. There are two main implications of their surveys that are relevant for this work. First, the two stars from HST photometry brighter than the TRGB may indeed belong to the AGB population of \andxviii\ and therefore may still be members (see Figure~\ref{fig:cmd3}).  Second, this dwarf galaxy must have faced star formation at some recent time, because bright AGB stars are at least intermediate in mass.  Their presence lends support to the younger ages found in the SFH \citep{mak17}.

\subsection{Kinematics}
\label{sec:resvel}
We measured the galaxy's kinematics by first assuming that $v_{\rm rot} = 0$ (Equation~(\ref{eq:6})). We used MCMC to find the most likely values of $\langle v_{\rm helio} \rangle$ and $\sigma_v$. After $<10$ MCMC trial runs for membership determination, with 5 walkers consisting of 1000 iterations each, we labeled 57 stars as probable velocity members of \andxviii\ (see Section~\ref{sec:members}).  

We evaluated the kinematic properties both including and excluding PAndAS9654, although the difference in the derived $\langle v_{\rm helio} \rangle$ is 0.1 km s$^{-1}$: $-337.3$~km~s$^{-1}$ with PAndAS9654 and $-337.2$~km~s$^{-1}$ without. Then, we calculated the two parameters in 10$^{5}$ iterations. For the rest of this paper, we quote values derived with the exclusion of PAndAS9654 (56 members).

Next we tried to derive whether \andxviii\ rotates by maximizing the likelihood in Equation~(\ref{eq:7}) with MCMC\@. In this case, 10$^{6}$ iterations converged to nearly the same $\langle v_{\rm helio} \rangle$ and $\sigma_v$ as the previous trial. The derived rotation velocity is significantly lower than the velocity dispersion (see Table~\ref{tab:results}). This means that \andxviii\ is a dispersion-supported dwarf galaxy (it does not exhibit significant rotation).

We estimated the dynamical mass within the half-light radius, $\log(M_{1/2}/M_\mathSun)$, using $M_{1/2}=3\sigma^2_{\rm los}r_{1/2}G^{-1}$ from \citet{wol10} and the half-light radius from \citet{mcc08} (see Table~\ref{tab:comparison}). The use of this equation is justified because \andxviii\ is a dispersion-supported galaxy.

\input{4_comparison.tab}

\subsection{Metallicity}
\label{sec:resabund}

We measured a mean metallicity of {\rm $\langle{\rm [Fe/H]}\rangle=-1.49$} and a dispersion of {\rm $\sigma_{\rm [Fe/H]}=0.36$} by fitting a Gaussian to the metallicity distribution. The luminosity--metallicity relation \citep{kir13} predicts that the mean metallicity of \andxviii\ would be {\rm ${\rm [Fe/H]} = -1.65 \pm 0.16$}, using the luminosity computed by \citet{mak17}.  This prediction is consistent at $1\sigma$ with our measurement of {\rm $\langle{\rm [Fe/H]}\rangle=-1.49$}.

Although we measured reliable abundances (less than 0.3~dex random uncertainties) for only \membersfeh\ member stars (and PAndAS9654), they show a clear metallicity gradient (Figures~\ref{fig:gradient_ra} and \ref{fig:gradient_dist}). Using the least-squares method, for 37 \andxviii\ members excluding PAndAS8654 and the most distant star, we measured:
\begin{equation}
\nabla_{\rm [Fe/H]}(r/r_{1/2})=-0.23\pm0.03~{\rm dex}~ r_{1/2}^{-1}\\
\label{eq:9}
\end{equation}

\begin{figure*}
\centering
\includegraphics[width=0.48\linewidth]{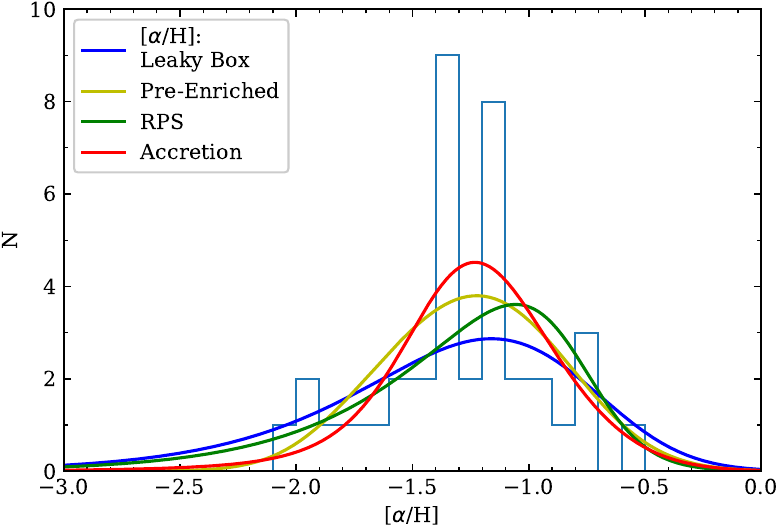}
\includegraphics[width=0.48\linewidth]{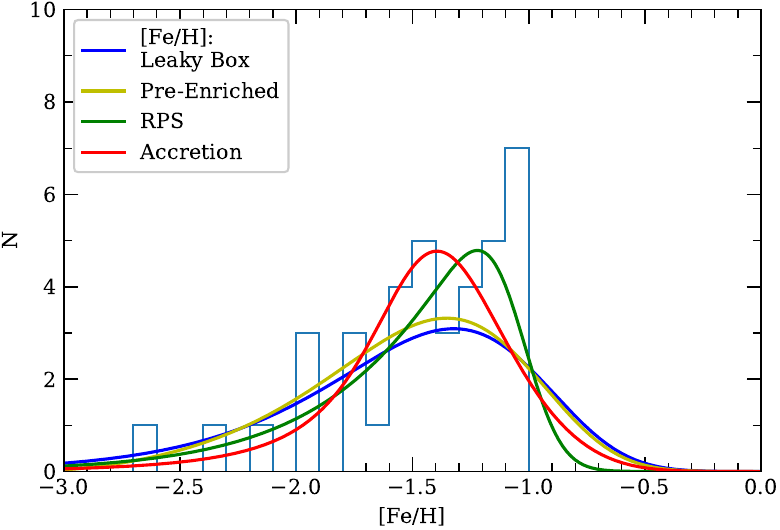}
\caption{Metallicity distributions for [$\alpha$/H] ({\it left}) and [Fe/H] ({\it right}) in \andxviii\@. The metal-rich end of the [Fe/H] distribution appears to be sharply truncated, but that truncation is less apparent in [$\alpha$/H]\@.  The best-fit chemical evolution models, convolved with the error kernel of the measurements, are shown as colored curves.}
\label{fig:chem}
\end{figure*}
\begin{figure}
\centering
\includegraphics[width=1.\linewidth]{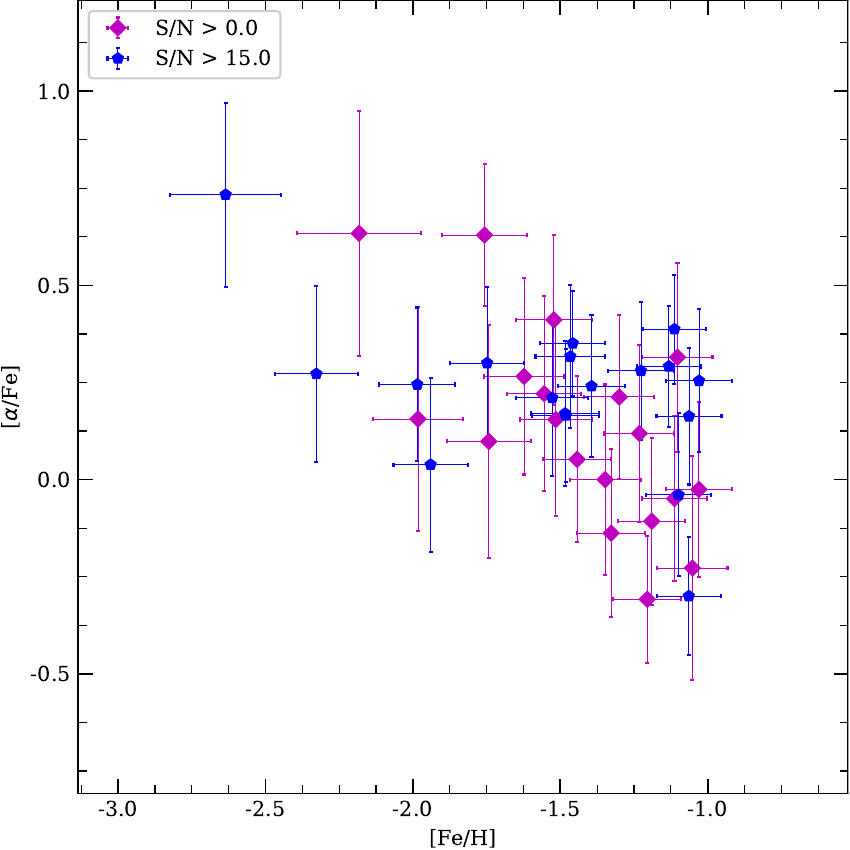}
\caption{Tinsley diagram (\alphafe\ vs.\ \feh) for 38 stars, excluding PAndAS9654. Stars with S/N {\rm $>$} 15 are shown in {\it blue}; lower S/N are shown in {\it purple}.}
\label{fig:abunds}
\end{figure}

\noindent
for the linear metallicity profile approximation.  We also re-calculated the gradient, excluding the next two outermost stars to examine whether they heavily influence the measured gradient:

\begin{equation}
\nabla_{\rm [Fe/H]}(r/r_{1/2})=-0.29\pm0.04~{\rm dex}~ r_{1/2}^{-1}
\label{eq:9b}
\end{equation}

The uncertainties on these intervals are calculated from the Python function \texttt{scipy.optimize.curve\_fit} \citep{2020SciPy-NMeth}.  The errors on the gradient do not include the negligibly small uncertainties on the distance from the center, $r$.

\citet{lea13} and \citet{tai22}, in their dwarf galaxy surveys, emphasized the importance of angular momenta, stellar mass, internal feedback (SFH), and galaxy interactions for the formation of metallicity gradients. The relationship and influence of these parameters remain obscure. The main \citeauthor{tai22}\ result is that the biggest metallicity gradient could be caused by dwarf-dwarf mergers.  Possible galaxies for which this happened include And~II, Phoenix, Sextans, Fornax, and NGC 6822 \citep[][and references therein]{tai22}. The steep spatial metallicity gradient of \andxviii\ seems to be in accordance with this set of dwarfs. \citeauthor{lea13} found that dIrrs often have shallow metallicity gradients, whereas dSphs have larger gradients. \andxviii\ conforms to this distinction.  However, in a larger set of dwarf galaxies, \citeauthor{tai22}\ could not corroborate the relation between dwarf galaxy morphology and strength of metallcity gradient. We discuss some further \andxviii's metallicity properties in the next sections and in Figures~\ref{fig:chem} and \ref{fig:abunds}.

\subsection{Detailed Abundance Ratios}

The Tinsley (\alphafe--\feh) diagram provides details about the SFH of \andxviii\ (Figure~\ref{fig:abunds}). The {\rm $[\alpha$/H]} of the galaxy increases due to core collapse (Type~II) supernovae, while \feh\ increases due to both Type~II SNe and thermonuclear (Type~Ia) supernovae.  The Type~Ia SNe explode only after a time delay of one Gyr -- hundreds of Myr. A shallower slope of the \alphafe--\feh\ relation indicates that both types of SNe were exploding at similar rates, which means  relatively continuous star formation. The \alphafe\ knee indicates when Type~Ia SNe dominated over Type~II SNe.  A steep decline in \alphafe\ reflects a period of waning star formation, when Type~II SNe were less frequent, but Type~Ia SNe were continuing to explode.  The slope of \alphafe\ for ${\rm [Fe/H]} > -1.5$ indicates that the star formation rate was steeply declining during this period of the galaxy's life.

\subsection{Chemical Evolution Models}
\label{sec:reschem}

In Section~\ref{sec:CEM}, we described analytic chemical evolution models of galaxy metallicity distributions under the assumption of instantaneous recycling. While this is a good assumption for $\alpha$ elements, \feh\ is created copiously by Type~Ia SNe, and thus iron is a delayed element that violates the instantaneous recycling approximation. On the other hand, the errors for {\rm $[\alpha$/H]}, which are derived as

\noindent
\begin{equation}
\delta [\alpha/{\rm H}]_{i} = \sqrt{\delta [\alpha/{\rm Fe}]_i^2 + \delta {\rm [Fe/H]}_i^2 + 0.106^2}
\end{equation}
\noindent
(where 0.106 is the systematic error on [$\alpha$/H], \citealt{kir10}), are bigger than the \feh\ errors, so fits to the \feh\ distribution are more reliable. To compare the models, we used the AICc:

\begin{equation}
{\rm AICc} = -2 \ln L + 2r + \frac{2r(r+1)}{N-r-1}
\end{equation}
where $L$ is the likelihood, $r$ is the number of model parameters, and $N$ is the number of stars. The smaller the AICc, the better the model.

The parameters derived from the \feh\ distribution (Figure~\ref{fig:chem}, right) for the Leaky Box and Pre-Enriched models are qualitatively similar for these models. The Accretion model well describes the overall shape of the distribution, giving \andxviii's current mass in units of its initial mass as $M\approx3.4$. The Accretion model is the most appropriate of these three models, according to its AICc, so we conclude that \andxviii\ experienced gas infall during its star formation lifetime. That infalling gas could be the return of gas after tidal stirring, a wet merger with another dwarf galaxy, or accretion of gas from the intergalactic medium (IGM)\@.

All three of these models fail to account for the sharp metal-rich cutoff at ${\rm \feh}\approx-1$. The Ram Pressure Stripping (RPS) model seems to solve this problem. Comparison of the best-fit stripping intensity ($\zeta \approx 4.6$) with some other dwarf galaxies \citep{kir13} indicates that \andxviii\ is among the most ram-pressure-stripped galaxies known. Furthermore, the best model is RPS, according to its AICc (Table~\ref{tab:mcmcwalkers}). 

The {\rm [$\alpha$/H]} histogram for \andxviii\ exhibits a rather different picture (Figure~\ref{fig:chem}, left). Both the Pre-Enriched and Accretion models seem to describe the [$\alpha$/H] distribution better than the RPS one. The RPS model is not required to explain the [$\alpha$/H] distribution due to the presence of a metal-rich tail. As noted by \citet{lea13}, the preference of the Pre-Enriched model can be caused by the undersampling of the metal-poor tail of the stars, although the \feh\ distribution contains a metal-poor tail. Our conclusions might be affected here by the spatial selection function \andxviii's members, coupled with its metallicity gradient (Figure~\ref{fig:gradient_dist}), which is more important than the sample size bias, as also shown by \citet{lea13}.

Thus, we have arrived at different conclusions depending on whether we treated metallicity as Fe or $\alpha$. However, the smaller errors on [Fe/H] make the \feh\ results more significant, so we cannot rule out that \andxviii\ faced ram pressure stripping in its past.


\section{Discussion}
\label{sec:rev}

In this section, we consider the evidence in support of or against three hypotheses for the origin of the dSph nature of \andxviii: the backsplash hypothesis, the merger hypothesis, and the self-quenching hypothesis.  This evidence comes from the above results, as well as some previous references, some of which are listed in Table~\ref{tab:comparison}.  For example, for the backsplash hypothesis, we consider whether the existing observations of the kinematics and chemical evolution of \andxviii\ support the hypothesis that it once passed through the virial radius of M31 \citep[which is more than 200 kpc,][]{gil12, tol12}. 

To emphasise the principal difference of the last hypothesis, it is important to mention that we do not have direct evidence whether the cause of this gas loss was an internal process, like feedback, or an external process, like stripping by M31. Internal processes, like gas stabilization after prolonged star formation or internal feedback (i.e., from supernova explosions and winds from low-mass stars) are not expected to be the main drivers of gas removal in dwarf galaxies \citep[][and references therein]{wei11}. \citeauthor{wei11} found that SFHs in dIrrs and dSphs are distinct in the recent past, especially within 1--2 Gyr. \andxviii's comparatively continuous star formation for several Gyr prior to ceasing 1.5 Gyr ago \citep{mak17} might indicate that external processes, such as ram pressure stripping or tidal stirring, completely removed all of \andxviii's star-forming gas in the recent past. However, internal processes are known from simulations and observations of the oldest stellar populations to be most important early in the Universe. DSphs, most of which formed the majority of their stars more than 10 Gyr ago, seem to lose their gas due to intensive Type~II supernova feedback \citep[e.g.,][]{ber18}. Deeper photometry of \andxviii\ that reaches the main-sequence turn-off of old populations can also improve the precision of the SFH at old times.  From those future observations, we can better infer the role of internal processes on gas loss at early times.

We also emphasize the importance of membership determination in our conclusions.  Membership in \andxviii\ is especially complicated due to the galaxy's distance and possible confusion with M31 halo stars.  All of our conclusions are predicated on the assumption that our sample contains very few nonmembers.

\subsection{Backsplash Hypothesis}

\subsubsection{Stellar Population}

If \andxviii\ is a backsplash galaxy, it could have experienced ram pressure stripping during its pericentric flyby.  In this context, it is useful to know the amount of remaining gas.  \citet{spe14} estimated an \ion{H}{1} mass upper limit of $4.8 \times 10^6~M_{\sun}$ for a distance of 1.21~Mpc \citep{mcc12}.  Corrected to 1.33~Mpc, the limit would be $5.8 \times 10^6~M_{\sun}$.  The nondetection of \ion{H}{1} is consistent with ram pressure stripping, but the upper limit is too large to be definitive (Appendix~\ref{sec:apprps}).

However, the SFHs determined by \citet{mak17} and \citet{wei19} using different CMD models might support ram pressure stripping in the past.  They used HST photometry, although it was not deep enough to reach the main-sequence turn-off of the oldest stars. For that reason, both works mentioned the low precision for the 12--14 Gyr period. \citet{mak17} discussed in detail the presence of old and intermediate populations in \andxviii\ based on the oblong red clump shape. The periods of star formation are 12--14 Gyr ago and about 1.5--8 Gyr ago with a starburst 8 Gyr ago and full quenching at 1.5 Gyr ago. The authors explained such a recent truncation as natural gas loss (due to supernova winds, for example) without the need for an external mechanism of gas loss. Despite that, the metallicity distribution of \andxviii\ has a sharp metal-rich truncation (Figure~\ref{fig:chem}).  As a result, the metal-poor tail appears consistent with a Leaky Box or Accretion model, which suggests that \andxviii\ was forming stars like a normal isolated dwarf galaxy, but then the sharp drop at ${\rm [Fe/H]} \sim -1$ is consistent with neither of these models. 

The \alphafe--\feh\ diagram also supports a complex SFH\@. The shallow slope of [$\alpha$/Fe] at ${\rm [Fe/H]} \lesssim -1.5$ is consistent with the analogous diagram for \andxviii\ obtained from coadded spectra by \citet{woj20}, although they have a smaller sample. As already mentioned (Section~\ref{sec:abund}), this shape of the curve indicates the relative frequency of Types~II and Ia SNe. When the galaxy had a metallicity ${\rm [Fe/H]} \lesssim -1.5$, the rates of the two types of SNe were comparable.  This confirms that the galaxy had a sufficient gas mass to power star formation for a while.  After that, at ${\rm [Fe/H]} \gtrsim -1.5$, the slope of \alphafe\ decreases steeply. Thus we conclude that the gas loss began in earnest when the metallicity was in the range $-1.5 < {\rm [Fe/H]} < -1.0$.

\subsubsection{Kinematics}

Now we consider \andxviii's orbit around M31, testing whether it is kinematically consistent with being a backsplash galaxy. We use \citeposs{tey12} definition of a backsplash galaxy:  objects on extreme orbits that have taken them through the inner $0.5~R_{\rm vir}$ of a larger potential and subsequently carried them back outside $R_{\rm vir}$. From cosmological Via Lactea II simulations, \citeauthor{tey12}\ estimated the fraction of this type of galaxy at about $13\%$ for the Local Group.  \citet{tol12} measured a radial velocity and a velocity dispersion of \andxviii\ from the SPLASH overview of M31 satellite system. They found that \andxviii\ is bound to M31 and possibly near its apocenter. Therefore, it might have once passed close to M31, qualifying it as a backsplash galaxy.

Starting from the two-body problem (\andxviii\ and M31), we can consider five observables: the distances between \andxviii, M31 and the MW, and the heliocentric radial velocities of \andxviii\ and M31. The three-dimensional distance between \andxviii\ and M31 is 579~kpc \citep{mak17}, and their relative velocity is $-36~{\rm km~s}^{-1}$ using an M31 heliocentric velocity of $-301$~km~s$^{-1}$ from \citet{kar06} (approximating the angle between the M31 line of sight and the \andxviii\ line of sight as zero). \andxviii's total orbital energy is negative if we assume that it has no tangential velocity component. For these assumptions, the highest tangential velocity to keep the system bound is about 154.5 km s$^{-1}$ relative to M31. For such a value, the proper motion relative to M31's proper motion will be about 24.5$~{\rm \mu as~yr}^{-1}$.  There are no current measurements of its proper motion.

Next, using the virial theorem and the energy conservation law, along with the observed kinetic energy per unit mass estimated from its radial velocity only, we obtain a semimajor axis of about 305~kpc. Then, with Kepler's laws and M31's mass of $\approx1.69 \times 10^{12}~M_{\rm \mathSun}$ \citep{mar12}, its orbital period is about 12~Gyr. The time since its closest passage through M31 is about 7.8~Gyr (see Appendix~\ref{sec:app2b})\@. Around the same lookback time, \andxviii\ experienced a spike in SFR \citep{mak17}. \citet{dic21} discussed how a pericentric passage could actually {\it increase} SFR, and \citet{miy20} found that observed dwarf galaxies reached their peak SFR at the time of infall (presumably right before their gas was stripped).  Nonetheless, \andxviii's minimum pericentric distance could be as small as 1.2 kpc, which is well within the range required for ram pressure stripping and significant tides. Appendix~\ref{sec:apprps} gives further details about the RPS dependence on the unknown tangential velocity, and in Appendix~\ref{sec:app2b} we consider the two-body modeling results for the backsplash hypothesis in greater detail than the rough discussion here.

\begin{figure*}
\centering
\includegraphics[width=0.3\linewidth]{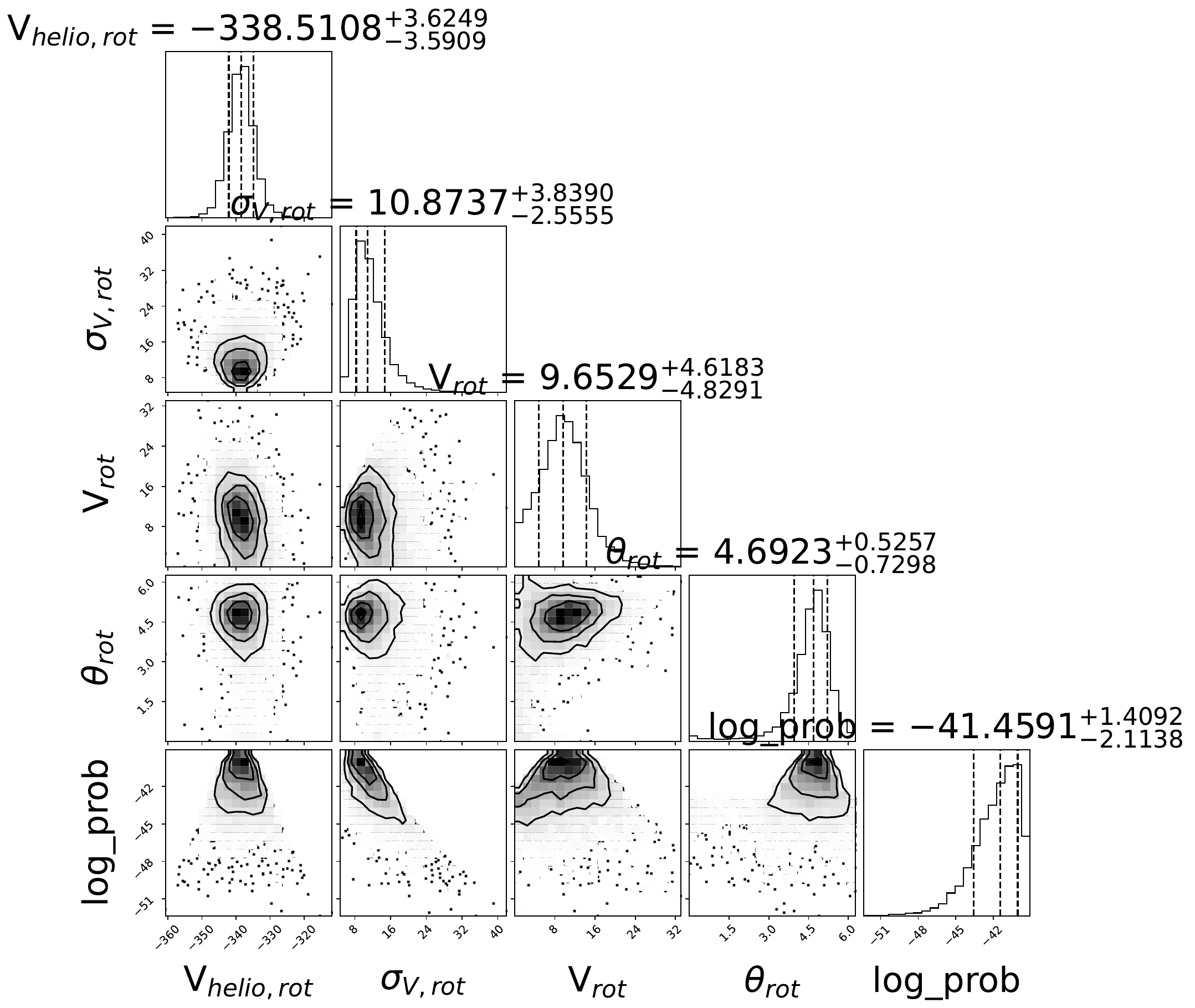}
\includegraphics[width=0.3\linewidth]{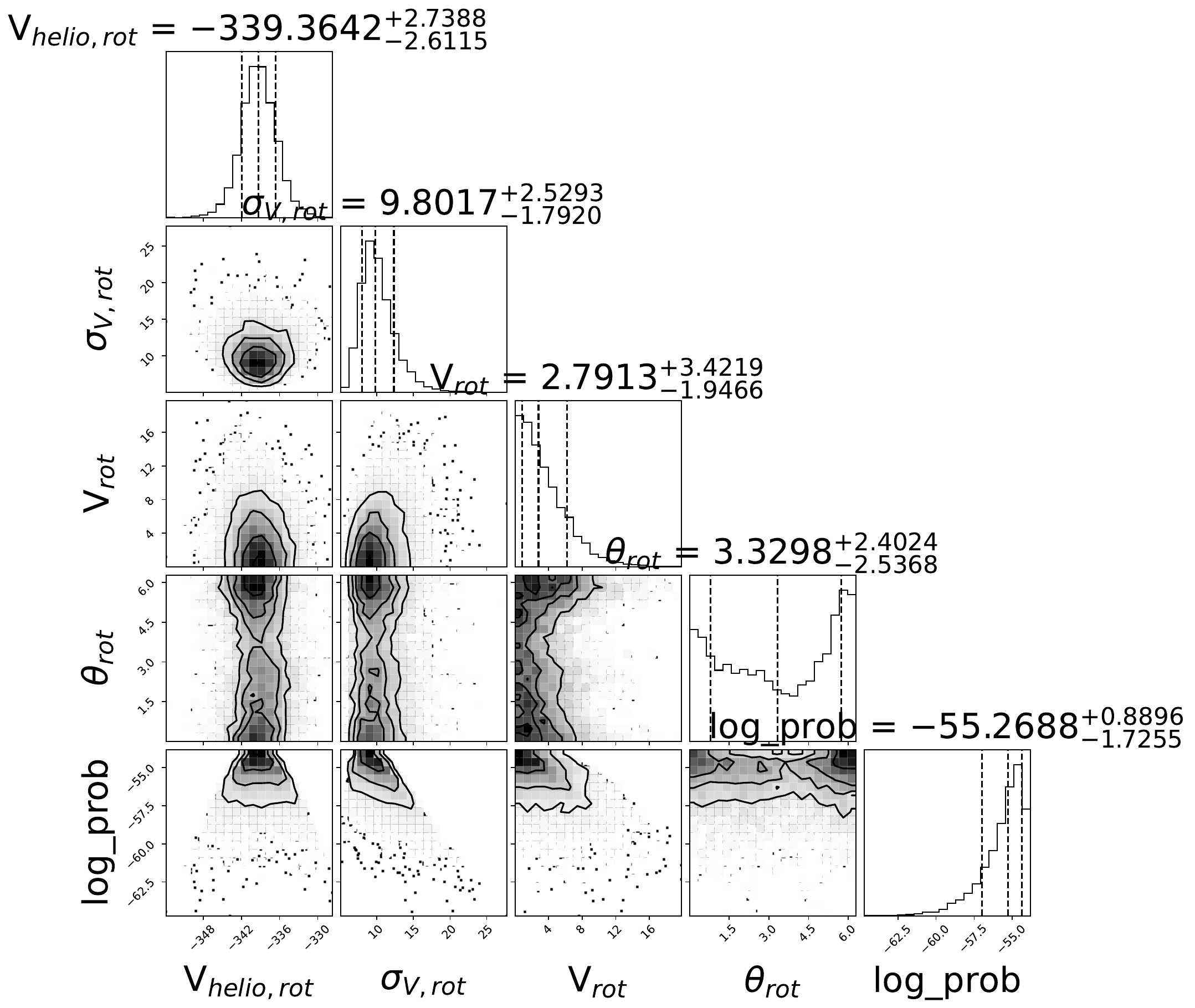}
\includegraphics[width=0.3\linewidth]{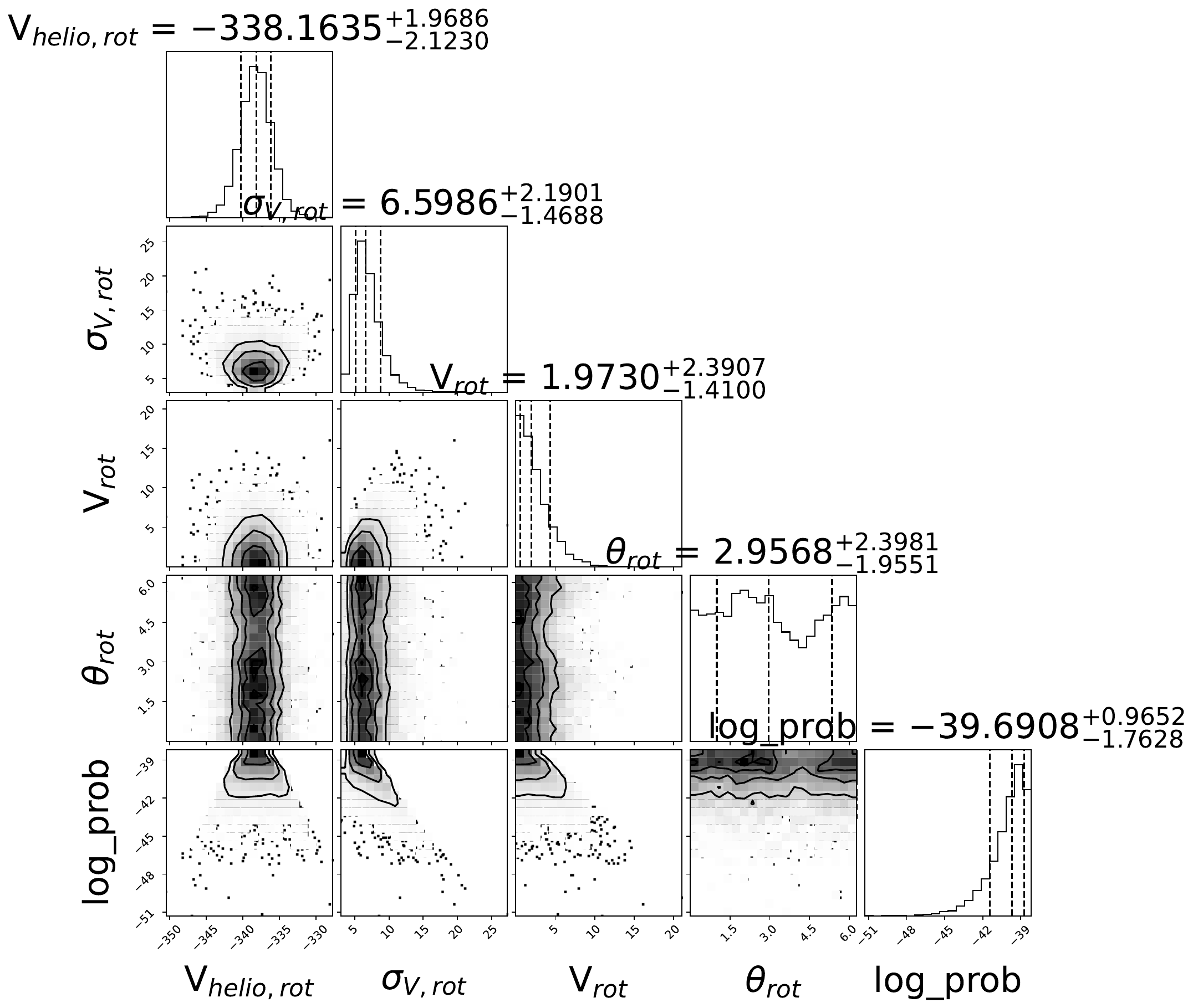}
\caption{{\it From left to right:} MCMC corner plots for rotational models of secure \andxviii\ members in the metallicity bins ${\rm [Fe/H]} <-1.6$,  $-1.6< {\rm [Fe/H]} <-1.25$, and ${\rm [Fe/H]} >-1.25$.  The parameters are mean heliocentric velocity ($v_{\rm helio,rot}$), velocity dispersion ($\sigma_{v,{\rm rot}}$), rotation velocity ($v_{\rm rot}$), and position angle ($\theta_{\rm rot}$).  The likelihood is ``prob''.}
\label{fig:corners}
\end{figure*}

We can also consider the three-body (\andxviii, M31, and the MW) problem without accounting for interaction between the MW and M31, i.e., assuming binary-like motion of the large galaxies.  The current distance between M31 and \andxviii\ (579~kpc) is comparable with the distance of M31 to the MW \citep[about 770 kpc,][]{mar12, kar06}. Taking into account the ratio of M31's mass to the MW -- about two to three -- the Roche lobe of M31 is expected to be more extended than the MW's, with a radius about 2/3 of the distance between the two galaxies. \citet{kne11} investigated renegade dwarf galaxies, which change their host galaxy. However, the dynamical modeling in Appendix~\ref{sec:app3b} exclude this possibility of \andxviii\ orbiting the MW in the past. Our three-body model indicates that \andxviii\ could have experienced a flyby of M31 at about 7--10 Gyr ago, depending on the unknown tangential velocity of \andxviii\ relative to M31's tangential velocity (see Appendix~\ref{sec:app3b})\@.

For simplicity, we did not consider cosmological effects.  With a Hubble constant of $H_0 = 68$~km~s$^{-1}$~Mpc$^{-1}$ \citep[c.f.,][]{desi24}, the current cosmological recession velocity of \andxviii\ is 90~km~s$^{-1}$ relative to the MW and 39~km~s$^{-1}$ relative to M31.  Therefore, there is a future opportunity to explore more sophisticated dynamical models.

In addition, the passage through a large host leads to tidal stirring or ram pressure stripping \citep{tey12}. From Figure~\ref{fig:obs} and \citet{hig21}, we can see that \andxviii\ has an oblong shape, which could be a sign of past tidal stripping \citep[see][]{ben18}. However, an increase in SFR, such as that inferred by \citet{mak17}, is not necessarily expected in these interactions.  On the other hand, interaction with a smaller (comparably sized) companion could increase the SFR \citep{ong09,lel14}.  The next section considers this possibility.


The most consequential assumption here is the neglect of the tangential component of \andxviii's orbital velocity. \andxviii\ appears behind M31 on its assumed orbit around M31, so the tangential component of its velocity could be bigger than the radial component, unless the orbital ellipticity is close to one. Some dwarf galaxies are on highly radial orbits \citep{sim18,li21}, so it is certainly possible that \andxviii\ is on such an orbit.  Nonetheless, we cannot definitely say that \andxviii\ is bound to M31 in the absence of tangential velocity.  Future observations with HST, JWST, Roman, or Euclid might be able to measure proper motion with the precision to rule out a bound orbit \citep[e.g.,][]{lib23}.

\subsection{Merger Hypothesis}
\label{sec:disc_kin}

\begin{figure}
\centering
\includegraphics[width=1.\linewidth]{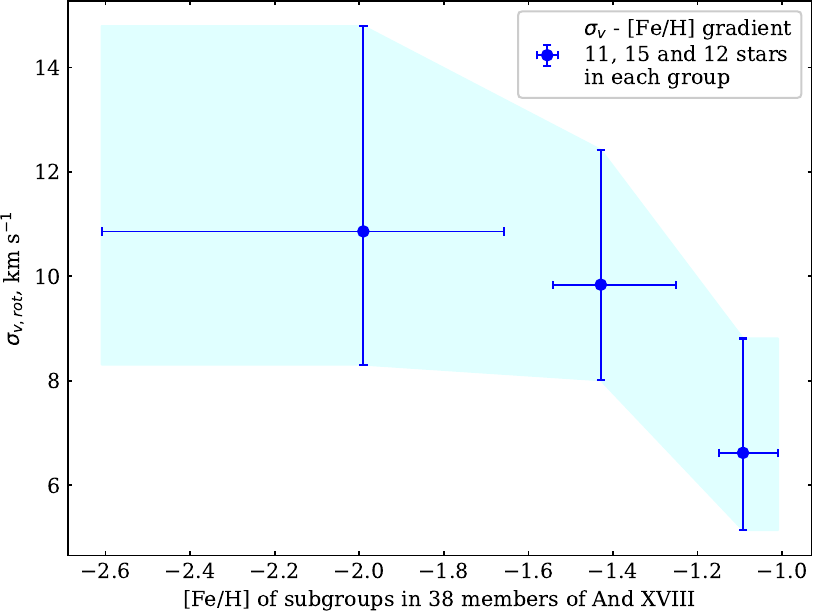}
\caption{Velocity dispersion estimated for secure \andxviii\ members in metallicity bins $\feh <-1.6$,  $-1.6< \feh <-1.25$, and $\feh >-1.25$ calculated with the rotational kinematical model. This result is in correspondence with the observed kinematic gradient in Figure~\ref{fig:gradient_dist} (right).}
\label{fig:sigmafeh}
\end{figure}

For the merger hypothesis, we can consider some inferences from \andxviii's internal kinematics. The signs of a major merger may include prolate rotation, starbursts, resonant ram pressure stripping, differences in the young and old stellar populations, and a steep metallicity gradient \citep{ong09,lel14,car21}. Some of these properties are present in \andxviii\@.  The galaxy did experience an increase in SFR 7.5~Gyr ago, followed by a period of continuous star formation \citep{mak17}. That continuation of star formation does not support the idea of a backsplash galaxy, wherein the interaction with a large galaxy is expected to remove all gas from the dwarf galaxy.  However, it could be consistent with a major merger.  Furthermore, \andxviii\ does have a steep metallicity gradient (Figure~\ref{fig:gradient_dist}), which is also true for the suspected merger remnants with prolate rotation \citep[And~II and Phoenix,][]{car21}.

We investigated kinematic differences between the older and younger populations, identified by \citet{mak17} as being split at $\feh = -1.6$ and $-1.25$.  Each group ($\feh <-1.6$,  $-1.6< \feh <-1.25$, $\feh >-1.25$) contains 10--15 stars.  Figure~\ref{fig:corners} presents the most likely parameters for the model with rotation (Equation~(\ref{eq:7})). Although all three groups are dispersion-supported, the dispersion for the two more metal-rich (younger) groups is lower, and the oldest stars are the most dynamically heated.  Evidence for rotation -- though not very significant -- is the strongest in the most metal-poor bin. The rotation axis position angle for the most metal-poor group is close to the semimajor axis defined by \citet{hig21}, who measured the galaxy's ellipticity to be about $0.4$ and the position angle to be 75.12$^{\circ}$ (4.45~rad, while 4.69~rad was obtained for the positional angle of the rotational axis for the most metal-poor stars); see Figure~\ref{fig:corners}. Thus, there is a possibility that \andxviii's older stars exhibit prolate rotation.

The merger of two gas-rich dwarf galaxies is theorized to induce centrally concentrated star formation with a velocity dispersion and radial extent smaller than the pre-merger population \citep{ben16}.  The merger drives gas to the center, where it is completely expended in new star formation and feedback-driven ejection \citep{tai22}. Figure~\ref{fig:gradient_ra} (right) and Figure~\ref{fig:sigmafeh} show that the metal-poor stars have a larger velocity dispersion than the metal-rich stars.  The latter plot was obtained after the application of the rotational kinematic model for metallicity subgroups in \andxviii\@. The comparatively dynamically hot metal-poor fraction of \andxviii\ stars may support the hypothesis that it originated in the merger of two roughly equal-sized dwarf galaxies \citep{wal11, gen19}.

\subsection{Self-quenching Hypothesis}

The measured SFH of \andxviii\ \citep{mak17} suggests extended star formation, even after the time of its possible close passage by M31.  It would be unusual for a dwarf galaxy of \andxviii's stellar mass to survive ram pressure stripping.  Therefore, it is possible that it was not ram pressure stripped or even a product of a major merger.  Instead, it could have quenched itself.

\citet{ari87} presented a galactic chemical evolution model wherein supernovae drive a wind from the galaxy.  For galaxies with a gas mass $\lesssim 4 \times 10^9~M_{\sun}$, the wind becomes a ``terminal wind'', which removes all gas from the galaxy and quenches it.  The resulting metallicity distributions can be sharply truncated at the metal-rich end, essentially because the wind interrupts the galaxy's otherwise closed-box chemical evolution.

Observed metallicity distributions of dwarf galaxies \citep[e.g.,][]{kir11a} rarely show such a sharp truncation.  This is especially true for isolated dwarf irregular galaxies \citep{kir13}.  Therefore, \andxviii\ would be unique among known isolated galaxies if it were self-quenched by a terminal wind.

\citeposs{mak17} SFH was observed from shallow HST images -- just one orbit per filter.  Therefore, the SFH was measured almost entirely from the red clump.  A much more precise SFH could be determined from photometry that reaches the old main-sequence turn-off.  Those observations would require much deeper space-based photometry, such as by many more orbits with HST or with a moderate investment of time on JWST\@.


\section{Conclusion}

We derived the kinematic and chemical properties of \andxviii\@. \andxviii\ is a dispersion-supported spheroidal galaxy. Sharp truncation in the [Fe/H] distribution might indicate sudden gas stripping, such as ram pressure during a close passage by M31.
The \alphafe--\feh\ diagram is consistent with the extended (but now quenched) SFH found by \citet{mak17}.  We considered three possible theories of \andxviii's origin:
\begin{enumerate}
\item \andxviii\ is a backsplash galaxy that once passed close to M31. With a small enough (yet unknown) tangential velocity, its radial velocity relative to M31 indicates that it is bound to M31. The position on the luminosity--metallicity diagram suggests past tidal interaction, which would have led to gas loss and would have truncated the SFH\@.
\item \andxviii\ is the result of the merger of two comparably sized dwarf galaxies. Its velocity dispersion gradient and metallicity gradient are well explained by this type of interaction.
\item Because isolated dwarf galaxies are rarely quenched, it is plausible that past environmental interaction rather than internal feedback quenched \andxviii\ \citep[e.g.,][]{wei11}.  Nonetheless, we cannot rule out that \andxviii\ quenched itself without stronger evidence of past environmental interactions.
\end{enumerate}

The observed orbital energy of \andxviii\ is consistent with the backsplash hypothesis. Alternatively, the SFH and metallicity gradient are consistent with the merger hypothesis. The inconsistency of both suggestions could mean that internal processes account for the expected gas loss, after all. Measuring a proper motion could disprove the backsplash hypothesis by proving that \andxviii\ is not bound to M31.

We conclude that the balance of the evidence slightly favors a backsplash origin to \andxviii, which is weakly associated (using the terminology of \citet{tey12}) with M31.  Its passage through M31 could have happened about 10~Gyr ago.  Instead of quenching all star formation as it passed through the host galaxy, it first intensified star formation and later lost its gas.

Overall, some of the evidence we have presented supports each of the above three hypotheses, and some of it is mildly inconsistent with each of them.  Additional data -- especially deeper, second-epoch space-based imaging taken about a decade after the first-epoch HST imaging -- will help resolve how \andxviii\ became an unusually quiescent yet isolated dwarf galaxy.

\begin{acknowledgments}
We thank L.\ Makarova for sharing the HST photometry catalog.  E.N.K.\ acknowledges support from NSF CAREER grant AST-2233781.
R.L.B. acknowledges support from NSF grants AST-0307842, AST-0607726, AST-1009882, and AST-1413269 for the SPLASH photometry and AST-2108616. Based in part on observations at NSF Kitt Peak National Observatory, NSF NOIRLab (NOIRLab Prop. ID  2010B-0596; PI: R.~Beaton), which is managed by the Association of Universities for Research in Astronomy (AURA) under a cooperative agreement with the U.S.\ National Science Foundation. The authors are honored to be permitted to conduct astronomical research on I'oligam Du'ag (Kitt Peak), a mountain with particular significance to the Tohono O'odham.
\end{acknowledgments}

{\it Facilities}: Keck:II(DEIMOS), Mayall (MOSAIC-1 wide-field camera).

{\it Software}: astropy \citep{ast13, ast18, ast22}, corner \citep{for16}, emcee \citep{for13}, matplotlib \citep{hun07}, numpy \citep{van11}, scipy \citep{vir20}, DAOPHOT \citep{1987PASP...99..191S}, DAOGROW \citep{1990PASP..102..932S}, ALLFRAME \citep{1994PASP..106..250S}.


\newpage
\begin{appendix}

In these appendices, we further investigate the possibility of a close passage of \andxviii\ by M31 and that flyby's hypothetical ability to truncate star formation.  The following sections consider the RPS criterion \citep{gg72}, possible orbits of \andxviii\ around M31 within a two-body interaction, and some possible orbits of \andxviii\ within a three-body problem. To simplify the orbital modeling, we will consider only those orbits that lie within the plane, which is derived by the current positions of MW, M31, and \andxviii\@. To make all terms as clear as possible, we also describe here the applied coordinate transformation.


\section{Coordinate Transformation}
\label{sec:coords}

M31's velocity within the equatorial system was taken as ($125.2,~-73.8,~-301.$)~km~s$^{-1}$; the distance of 770 kpc and the proper motion along R.A. and Dec. directions are from \citet{mar12}; and the radial velocity is from \citet{kar06}\footnote{The M31 center coordinates were taken from \url{https://esahubble.org/images/opo1204b/}}. The equatorial proper motion, radial velocity, and coordinates were transformed into the heliocentric galactic Cartesian system using the \texttt{astropy.coordinates} package. $X$ is directed to the MW center, $Y$ is directed along the solar rotational motion within the MW plane, and $Z$ is directed toward the MW north pole so that $XYZ$ is a right-handed system. By adding the solar velocity and subtracting the MW center coordinates, taken as ($9.5,~250.7,~8.56$)~km~s~$^{-1}$ and ($8.249,~0,-0.0208$)~kpc from \citet[][and references therein]{ak24}, we obtained the galactocentric coordinates and velocities within the reference system, which we refer to as $XYZ$.

Next, as the two-body problem can be simplified to two dimensions, we will use the $XYZ$ transformation to limit our consideration of all possible orbits for \andxviii\ to those that lie in the new $X'''Y'''$ plane (the aforementioned plane defined by the MW, M31, and \andxviii). To realize this, we applied first the rotation for $\phi_1$ with $Z=Z'$, second the rotation for $\phi_2$ with $X'=X''$, and third the rotation with $\phi_3$ and $Y''=Y'''$, where the rotation is counterclockwise. The A18 subscripts in the equations below are for \andxviii's values. The applied angles are as follows:

\begin{equation}
\phi_1 = \arctan \left( \frac {Y_{A18}}{X_{A18}}\right) - \frac{\pi}{2};~~~
\phi_2 = \arctan \left( \frac {Z_{A18}}{\sqrt{X_{A18}^2 + Y_{A18}^2}}\right);~~~
\phi_3 = 3\pi - \arctan \left( \frac {Z''_{M31}}{X''_{M31}}\right)
\end{equation}
with $arctan$ derived between $0$ and $2\pi$, where
\begin{gather}
\begin{bmatrix} X'' \\ Y'' \\ Z''\end{bmatrix}
=
   \begin{bmatrix}
   1        & 0              & 0             \\
   0        & \cos (\phi_2)  & \sin (\phi_2) \\
   0        &-\sin (\phi_2)  & \cos (\phi_2) \\
    \end{bmatrix} 
    \begin{bmatrix}
   \cos (\phi_1) & \sin (\phi_1) & 0 \\
  -\sin (\phi_1)  & \cos (\phi_1) & 0 \\
   0             &  0            & 1 \\
    \end{bmatrix} 
    \begin{bmatrix} X \\ Y \\ Z \end{bmatrix}
\end{gather}
which gives:
\begin{equation}
Z''_{M31} = \sin (\phi_2) \sin (\phi_1) X_{M31} - \sin (\phi_2) \cos (\phi_1) Y_{M31} + \cos (\phi_2) Z_{M31};~~~
X''_{M31} = \cos (\phi_1) X_{M31} + \sin (\phi_1) Y_{M31} 
\end{equation}

The overall transformation looks like:
\begin{gather}
\begin{bmatrix} X''' \\ Y''' \\ Z'''\end{bmatrix}
=
   \begin{bmatrix}
   \cos (\phi_3)  & 0 & -\sin (\phi_3) \\
   0              & 1 & 0              \\
   \sin (\phi_3)  & 0 &  \cos (\phi_3) \\
    \end{bmatrix} 
   \begin{bmatrix}
   1        & 0              & 0             \\
   0        & \cos (\phi_2)  & \sin (\phi_2) \\
   0        &-\sin (\phi_2)  & \cos (\phi_2) \\
    \end{bmatrix} 
    \begin{bmatrix}
   \cos (\phi_1) & \sin (\phi_1) & 0 \\
  -\sin (\phi_1)  & \cos (\phi_1) & 0 \\
   0             &  0            & 1 \\
    \end{bmatrix} 
    \begin{bmatrix} X \\ Y \\ Z \end{bmatrix}
\end{gather}

The obtained transformation aligns the $Y'''$ axis with the line that passes through the MW center and \andxviii, and it sets the plane determined by the MW center, M31, and \andxviii\ to be coincident with $X'''Y'''$ plane. The new coordinate system center is at the MW center (i.e., only rotations were applied).

\begin{figure*}
\centering
\includegraphics[width=.48\linewidth]{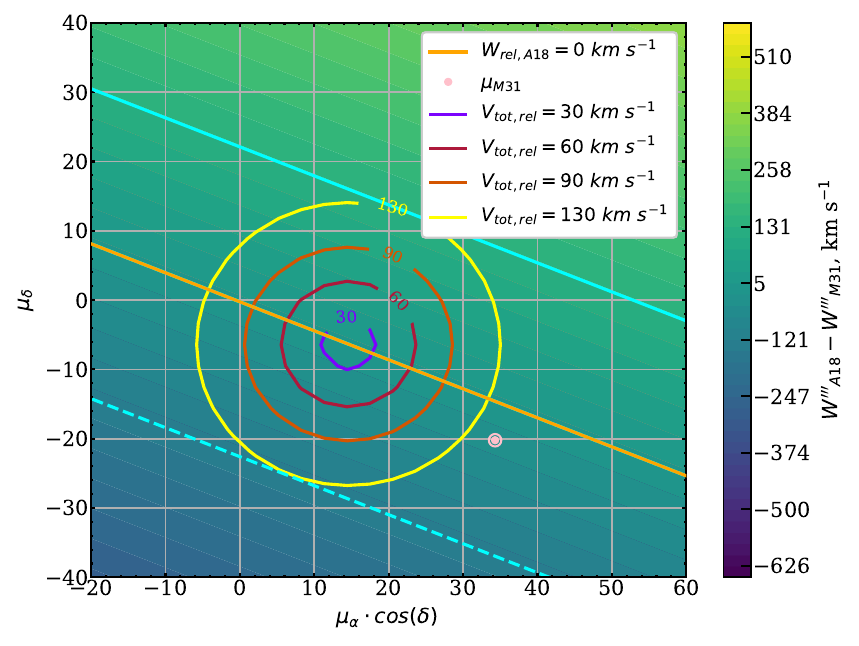}
\includegraphics[width=.48\linewidth]{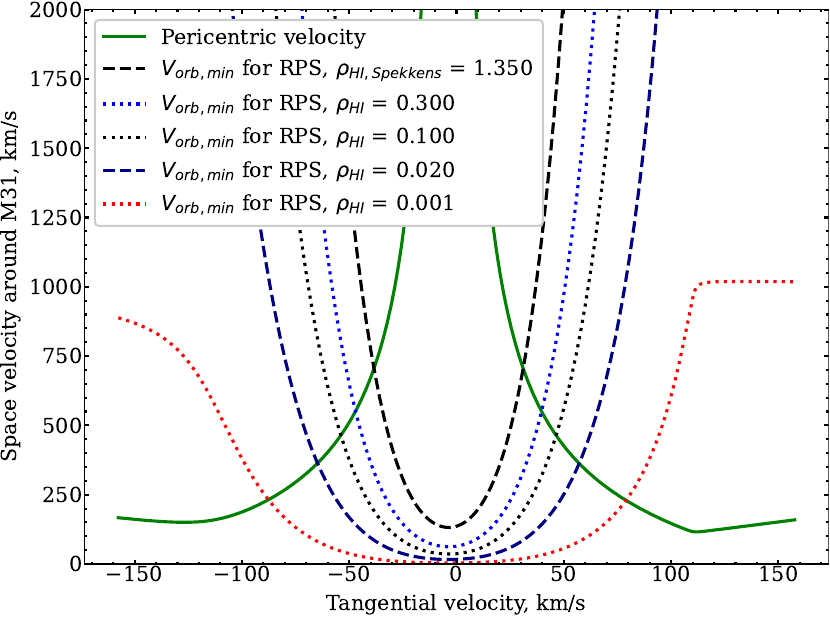}
\caption{{\it Left:} Possible $W'''_{A18}-W'''_{M31}$ values for a selected range of $\mu_{\alpha}\cos(\delta),~\mu_{\delta}$ (in $\mu$as~yr$^{-1}$). The {\it orange} line is for $W'''_{A18}-W'''_{M31}=0$, and the {\it blue} lines are for $130$~km~s~$^{-1}$. The {\it circles} show \andxviii's total velocity relative to M31. These regions are overlapping with the $V'''_{A18}-V'''_{M31}$ values from Sections~\ref{sec:app2b} and \ref{sec:app3b}. The {\it pink} point shows the adopted proper motion of M31 \citep{mar12}. {\it Right:} Pericentric velocity and lowest orbital velocity for various hydrogen densities, based on the ram pressure stripping condition from \citet{gg72} (Section~\ref{sec:apprps}). For absolute values of relative tangential velocities (see Section~\ref{sec:coords}), if the velocity is larger than 31 -- 38~km~s$^{-1}$, the pericentric distance will be far enough and the pericentric velocity will be small enough for \andxviii\ $not$ to have experienced the RPS in the past. Therefore, RPS is ineffective where the green curve lies below the RPS curve for the various hydrogen densities.  The M31 halo and \andxviii\ ISM gas densities are in ${\rm cm}^{-3}$.}
\label{fig:pmra}
\end{figure*}

The obtained transformation is constrained in equatorial coordinates by the known radial velocity of \andxviii\ ($-337.2$~km~s$^{-1}$) and by our requirement that $W'''_{A18}-W'''_{M31}=0$~km~s~$^{-1}$ in the new coordinate system. Figure~\ref{fig:pmra} (left) is color-coded by the $W'''_{A18}-W'''_{M31}$ values, and the orange line indicates $W'''_{A18}-W'''_{M31}=0$~km~s~$^{-1}$. The colored circles illustrate the absolute values of \andxviii's velocity relative to M31.  For the part of the orange line on the diagram within the circles, $V'''_{A18}-V'''_{M31}$, the projection of \andxviii's relative velocity toward the MW center will be $-18.4\pm0.4$~km~s~$^{-1}$. The nonconstancy of the relative radial velocity ($V'''_{A18}-V'''_{M31}$) for different proper motions, for which $W'''_{A18}-W'''_{M31}=0$, is the motivation for transforming from equatorial coordinates to a new cooridnate system. However, the relative tangential component as $|U'''_{A18}-U'''_{M31}|\lesssim150$~km~s~$^{-1}$, which causes \andxviii\ to be bound to M31, also makes the variation in $V'''_{A18}-V'''_{M31}$ small ($\pm0.4$~km~s~$^{-1}$). Therefore, we will consider the relative radial velocity of \andxviii\ as a constant in our discussion.

To summarize, we take \andxviii's $v_{\rm r}$ velocity to be $V'''_{A18}-V'''_{M31}=-18.4$~km~s$^{-1}$. Here, $v_{\tau}$ (\andxviii's tangential velocity relative to M31) is equal to $U'''_{A18}-U'''_{M31}$. We use the subscript $0$ to assign \andxviii's orbital parameters relative to M31 at the present time.  In these appendices, the terms ``relative line-of-sight velocity'' and ``relative radial velocity'' are projected onto the line connecting the MW center to \andxviii, not the line connecting the Sun to \andxviii\@. Thus, $v_{r0}$ and $v_{\tau 0}$ are defined as \andxviii's radial and transverse velocities relative to the MW center at the current time, and the following equations describe \andxviii's velocity relative to M31:

\begin{equation}
U'''_{A18}-U'''_{M31}=v_{\tau 0};~~~V'''_{A18}-V'''_{M31}= v_{r0} = -18.4~km~s^{-1};~~~W'''_{A18}-W'''_{M31}=0~km~s^{-1}
\end{equation}

\section{RPS after M31 Halo Passage}
\label{sec:apprps}

We consider whether M31 could have gradually tidally stripped \andxviii\ without engaging in rapid RPS\@.  In this scenario, the gas loss could be gradual enough to allow for the extended SFH observed by \citet{mak17}.

From \citet{spe14} (and references therein, including \citealt{gg72}), the condition for RPS is

\begin{equation}\begin{split}
\rho_{\rm M31} v_{\rm orb}^2 > 5 \rho_{\rm dw} \sigma_{\rm dw}^2
\end{split}\end{equation}
where $\rho_{\rm M31}$ is the gas density of the host galaxy halo, $v_{\rm orb}$ is the dwarf galaxy's orbital velocity at the corresponding $\rho_{\rm M31}$, $\rho_{\rm dw}$ is the ISM gas density of the dwarf galaxy, and $\sigma_{\rm dw}$ is the velocity dispersion of the dwarf galaxy's stars.

If we expect \andxviii\ to be bound to M31 and to be gradually tidally stripped, but not ram pressure stripped, its pericentric velocity should be $low~enough$ (Figure~\ref{fig:pmra}, right). For the two-body problem (Appendix~\ref{sec:app2b}), the pericentric velocity depends on the unknown current tangential velocity relative to M31, but the hydrogen density in M31's halo will also depend on the pericentric distance. The particle density may be comparable with a hot electron density within a halo, taken from \citet{tah22} as:

\begin{equation}\begin{split}
n_e^{\rm H}(r) = \frac{n_0^{\rm H}}{\mu_e\left(r/r_c+0.75\right)\left(r/r_c+1\right)^2}
\end{split}\end{equation}

\noindent where $n_0^{\rm H}=3.4 \times 10^{-2}~{\rm cm}^{-3}$, $r_c=15$~kpc, and $\mu_e=1.18$. Taken all together, the constraints on the relative tangential velocity of \andxviii\ to make it bound with M31 (details in Appendix~\ref{sec:app2b}) and shielded from complete ram pressure stripping (with the dSph gas density at its maximum possible value, 1.35~cm$^{-3}$, \citealt{spe14}) are about

\begin{equation}
31~{\rm km~s}^{-1}< v_{\tau 0} <  157.6~{\rm km~s}^{-1}
\end{equation}


\section{Two-body Problem}
\label{sec:app2b}

We consider the orbit of \andxviii\ around M31, assuming a point-like, static potential for both galaxies.  The results are illustrated in Figure~\ref{fig:orbit}. In Appendix~\ref{sec:app3b}, we extend the model to include the MW\@.

We used $M_{\rm M31}=1.69 \times 10^{12}~M_{\mathSun}$, $M_{\rm MW}=1.5 \times 10^{12}~M_{\mathSun}$, $v_{r, {\rm M31}}=-301~{\rm km~s}^{-1}$, $v_{r,{\rm AndXVIII}}=-337.2~{\rm km~s}^{-1}$, $r_0=579~{\rm kpc}$ as the 3D distance from \andxviii\ to M31, and $r_{p}=113.6~{\rm kpc}$ as the projected distance between M31 and \andxviii\ (the absolute value of the $X'''$ coordinate of the M31, see Section~\ref{sec:coords}; \citet{mak17, mar12}, and \citet{kar06}). The orbit is assumed to lie in the plane defined by the MW center -- \andxviii\ line and M31's position, so $v_{r0} \approx -18.4~{\rm km~s}^{-1}$ (see Section~\ref{sec:coords}).

\begin{figure*}
\centering
\includegraphics[width=.95\linewidth]{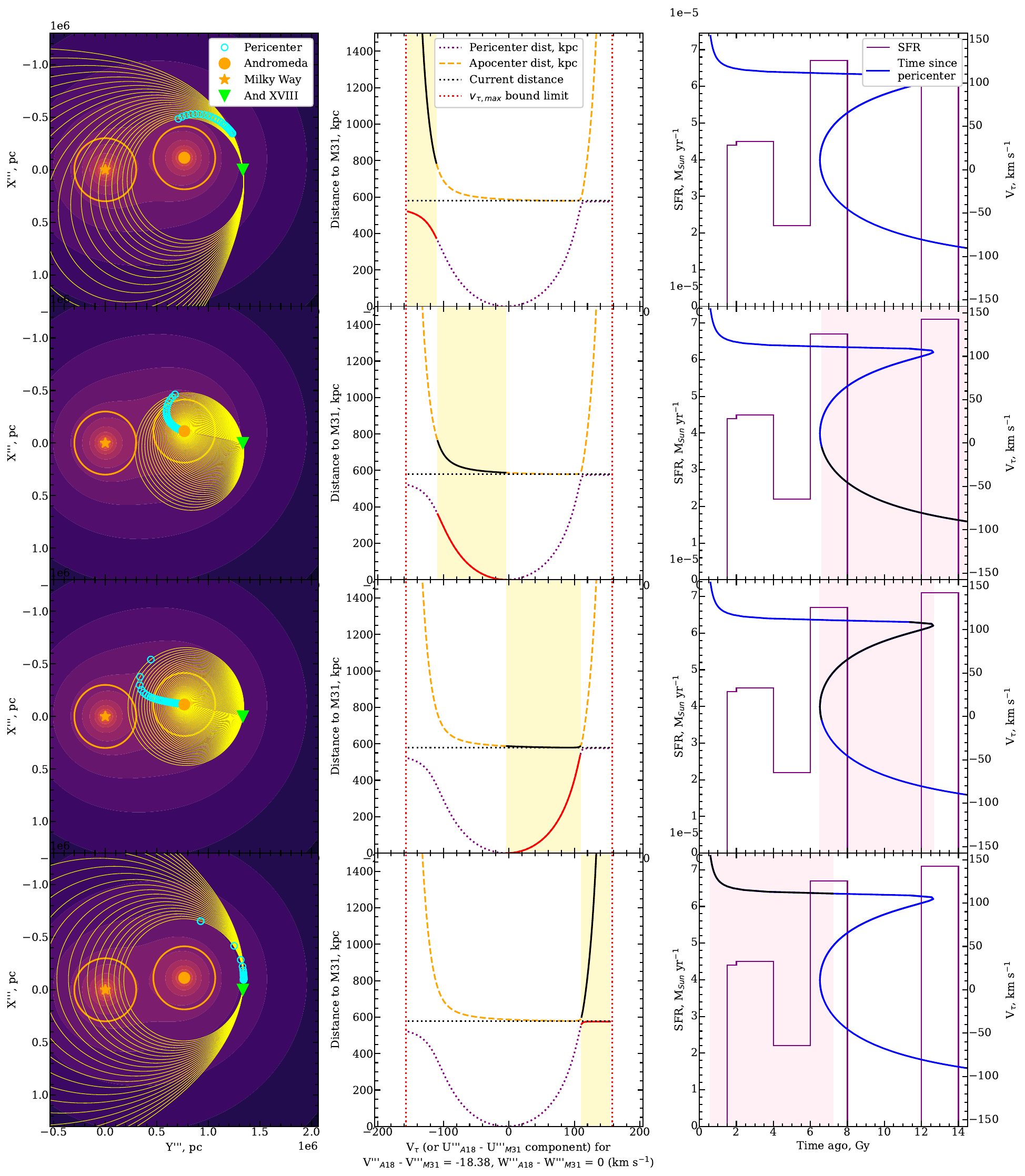}
\caption{{\it Left:} \andxviii's expected orbits for different tangential velocities.  The transverse motion is positive for the downwards direction. The colored background of MW and M31 gravitational potential is shown for the reference and was not used for this modeling. The different yellow orbits correspond to the different assumed tangential velocities, shadowed in the {\it middle {\rm and}  right} panels (not for different times). {\it Middle column:} Distances from M31 for varying values of $v_{\tau 0}$. {\it Right:} SFR \citep{mak17} and tangential velocity ($v_{\tau 0}$) vs.\ time since the last pericentric passage. The ``upper'' branch of the time of pericentric flyby ({\it right} panels) preclude RPS because those pericentric distances are outside of the virial radius of M31, illustrated by $cyan$ on the {\it left} panels. The second and the third rows give the region where RPS is possible.  The first and the last rows include orbits that intercept the virial radius of the MW\@.  For these situations, the two-body problem is insufficient (see Appendix~\ref{sec:app3b}).
}
\label{fig:orbit}
\end{figure*}

For the pericentric and apocentric positions, the distances and velocities can be found from energy and angular momentum conservation laws written as

\begin{equation}\begin{split}
\frac{(v_{r0}^2+v_{\tau 0}^2)}{2} - \frac{G M_{\rm M31}}{r_0} = \frac{v^2}{2} - \frac{G M_{\rm M31}}{r}\\
r_0 \Bigl[-v_{r0} \sin\left(\frac{r_p}{r_0}\right) + v_{\tau 0} \cos\left(\frac{r_p}{r_0}\right)\Bigr] = r v
\end{split}\end{equation}

The equations for distance and velocity at pericenter (+) and apocenter ($-$) can be obtained from

\begin{eqnarray}
r_{\pm} &=& \frac{-1 \pm \sqrt{1+4\Bigl[\left(-v_{r0} \sin\left(r_p/r_0\right) + v_{\tau 0} \cos\left(r_p/r_0\right)\right)^2 r_0/(2 G M_{\rm M31})\Bigr]\Bigl[\left(v_{r0}^2+v_{\tau 0}^2\right)r_0/(2 G M_{\rm M31})-1\Bigr]}}{(2/r_0)\Bigl[\left(v_{r0}^2+v_{\tau 0}^2\right)r_0/(2 G M_{\rm M31})-1\Bigr]}\\
v_{\pm} &=& \frac{r_0}{r_{\pm}}\left(-v_{r0} \sin\left(\frac{r_p}{r_0}\right) + v_{\tau 0} \cos\left(\frac{r_p}{r_0}\right)\right)
\end{eqnarray}

The upper limit for the unknown transverse velocity relative to M31 is obtained from the equality of energy to 0, which is around $157.6~{\rm km~s}^{-1}$. The apocentric and pericentric distances for zero tangential velocity will be about 586.4~kpc and 0.3~kpc respectively, with orbital velocities about 3.6~km~s$^{-1}$ and 6975~km~s$^{-1}$. Necessarily, this would mean that \andxviii\ lost all its gas due to ram pressure stripping (and tidal destruction).

\begin{figure*}
\centering
\includegraphics[width=0.49\linewidth]{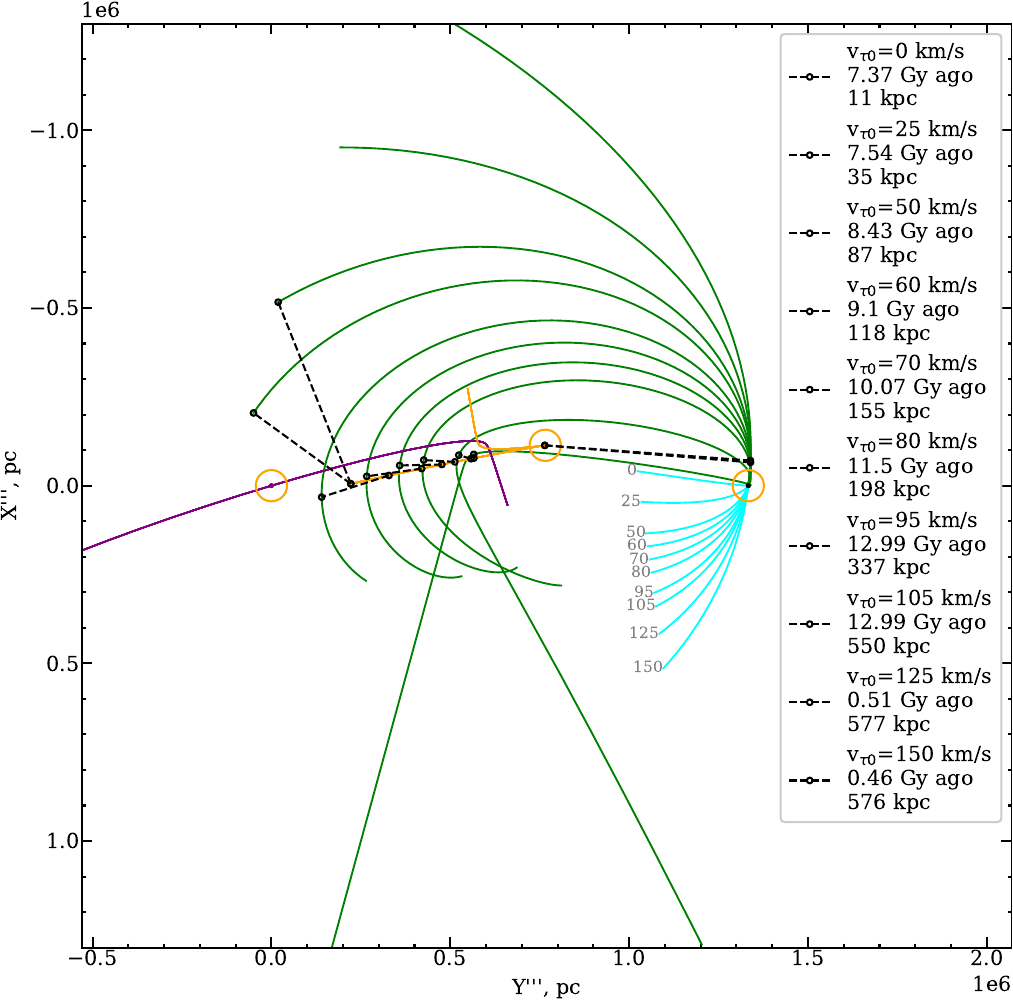}
\includegraphics[width=0.49\linewidth]{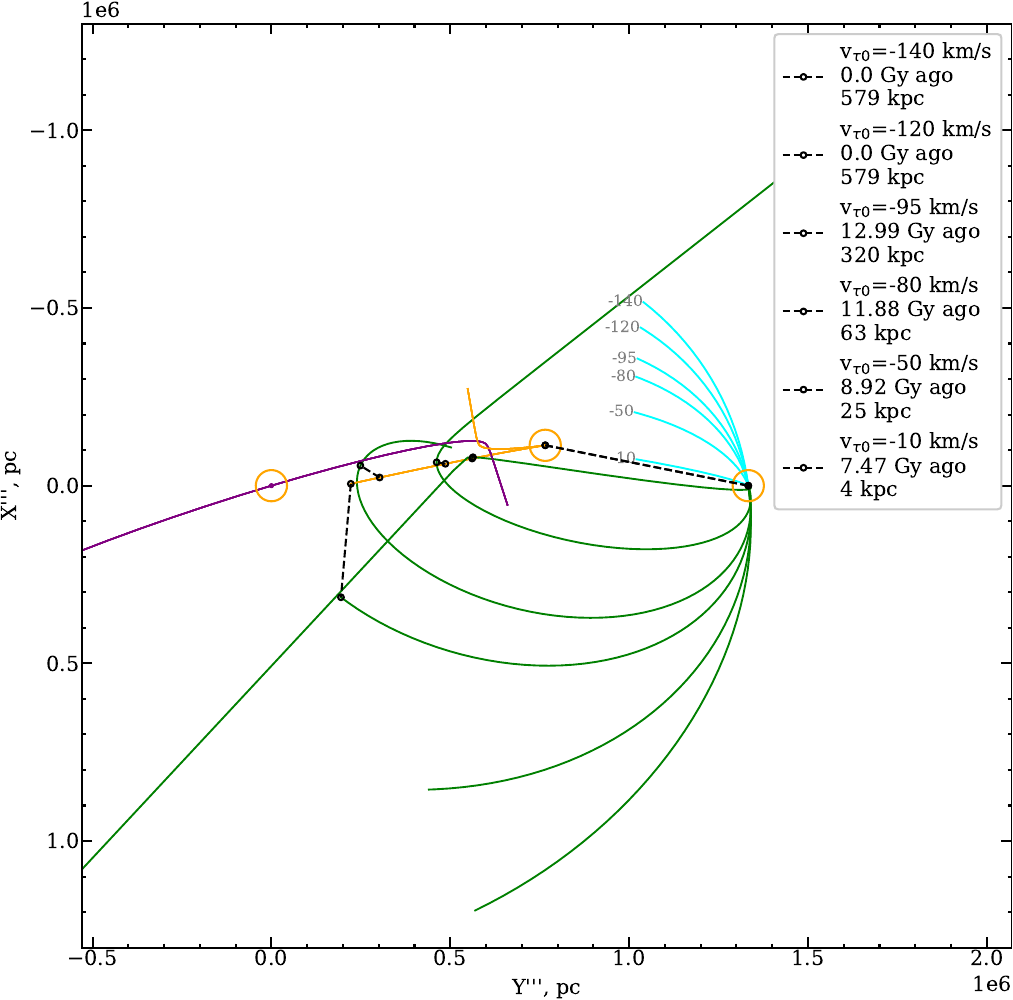}
\caption{{\it Left:} Three-body modeling for assumed v$_{\tau 0}$ of \andxviii\ relative to M31 as a positive value (clockwise). {\it Right:} The same for the negative relative to M31 tangential velocities (counterclockwise). {\it Green} color is for the past motion of \andxviii\ (up to 13 Gy ago), {\it blue} is for future motion (up to 3.86 Gy, before the expected M31--MW interaction, \citealt{mar12b}), {\it orange} is for M31 motion, with {\it orange} circles for current M31, MW, and \andxviii\ positions, {\it purple} is for MW motion; and {\it black dashed} segments indicate the shortest lines between \andxviii\ and M31, with {\rm v$_{\tau 0}$}, time of passage, and the distance as assigned in the plot's legend.}
\label{fig:body3}
\end{figure*}

Now we can consider the time since the last pericentric passage (Figure~\ref{fig:orbit}). The eccentricity dependence on the unknown tangential velocity is

\begin{equation}\begin{split}
e = \sqrt{1 + 4 \Bigl[-v_{r0} \sin \left(\frac{r_p}{r_0}\right)+v_{\tau 0} \cos \left(\frac{r_p}{r_0}\right)\Bigr]^2 \frac{r_0}{2 G M_{M31}} \Bigl[\frac{(v_{r0}^2+v_{\tau 0}^2) r_0}{2 G M_{M31}}-1\Bigr]}
\end{split}\end{equation}

The positional angle ($\theta$) is substituted into the time integral for this problem as
\begin{eqnarray}
\tan{\frac{\theta_0}{2}} &=& \sqrt{\frac{1+e}{1-e}} \tan{\frac{\psi_0}{2}} \\
r_0 &=& a(1-e\cos{\psi_0}) \\
\tan{\frac{\psi_0}{2}} &=& \sqrt{2 \Bigl[\frac{1}{e}\left(1-\frac{r_0}{a}\right)+1\Bigr]^{-1}-1}\\
t_0 &=& \frac{\Bigl[r_0 \Bigl[-v_{r0} \sin\left(r_p/r_0\right) + v_{\tau 0} \cos\left(r_p/r_0\right)\Bigr]\Bigr]^3}{\Bigl[G M_{M31}\Bigr]^2} \int_0^{\theta_0}{\frac{d \theta}{\left(1+e \cos{\theta}\right)^2}}
\end{eqnarray}
where $r_0$ is the current distance from \andxviii\ to M31, $\theta_0$ is the current position angle of \andxviii\ on its orbit, and $\psi_0$ is the current eccentric position angle of \andxviii\@. The integration starts from 0, the position angle of the pericenter.

We considered the tangential velocity of \andxviii\ in both directions within the M31 -- \andxviii\ plane. The reader may compare Figure~\ref{fig:orbit} (the first and the last rows) with the Figure~\ref{fig:body3}.

To summarize the results illustrated in Figure~\ref{fig:orbit}, a natural limit on the pericentric distance to {\it experience} RPS can be related to the range of relative tangential velocity that leads to a semimajor axis smaller than the \andxviii's current distance to M31: {\rm $|v_{\tau 0}|<110.5$} km {\rm s$^{-1}$}. This situation is shown in the second and third rows of Figure~\ref{fig:orbit}, where the pericentric points (cyan) are preferentially inside the virial radius of M31.


\section{Three-body Problem}
\label{sec:app3b}

We show some of \andxviii's possible orbits in the three-body problem including \andxviii, M31, and the MW (Figure \ref{fig:body3}\footnote{\url{https://github.com/blbadger/threebody}}).  The time step is 0.01 Gy. We assumed static, point-like potentials for each galaxy, and we ignored cosmological expansion. The variables are defined within the reference system shown in Figure~\ref{fig:body3}, with the MW at the center at the current time ($z = 0$). M31's current velocity is defined to be 0, and we adopted the velocity of the MW relative to M31 as M31's velocity with the opposite sign (described in Appendix~\ref{sec:coords}). \andxviii's mass was assigned as $10^{7.38}~M_{\mathSun}$ (Table~\ref{tab:comparison}). 

Both panels of Figure~\ref{fig:body3} are in accordance with Figure~\ref{fig:orbit}, although the $X'''Z'''$, $Y'''Z'''$ projections have not been described here.

The closest distances from \andxviii\ to M31 ({\it black dashed} lines in Figure~\ref{fig:body3}) show that the upper limit on $v_{\tau 0}$ to experience RPS is about 80 km~s$^{-1}$ for the positive direction (left panel) and about $-90$~km~s$^{-1}$ for the negative direction (right panel).  These tangential velocities correspond to pericentric distances of $r_{\rm min}>r_{\rm vir,M31}\approx 200$~kpc \citep{gil12, tol12} for $|v_{\tau 0}|> 80~(90)$~km~s$^{-1}$.  These numbers seems to agree with our findings from the two-body problem (Appendix~\ref{sec:app2b}).

Our conclusions about the backsplash hypothesis, considering the three-body problem, are as follows:
\begin{enumerate}
\item \andxviii\ cannot be a renegade dwarf galaxy. If it is a backsplash dSph, it passed near M31, not the MW\@.
\item It is possible to consider proper motion in both the clockwise and counterclockwise directions in the LOS--M31 plane. RPS is possible in either scenario.
\item The two-body problem seems to be sufficient here because the MW is a minor influencer of \andxviii's orbit. In other words, the tangential velocity limits derived in Section~\ref{sec:app2b} can be applied to all possible orbit inclinations, not only for the plane of motion as defined in Appendix~\ref{sec:coords}.
\end{enumerate}

\end{appendix}

\newpage
\bibliography{andromeda18}
\bibliographystyle{apj}

\end{document}

%% file: and18_10_objects_main.tab
\begin{rotatetable*}
\begin{deluxetable*}{ccccccccccccccccccc}
\label{tab:mrt}
\centerwidetable
\tabletypesize{\small}
\tablewidth{0pt}
\tablecolumns{19}
\tablecaption{Catalog of Velocities and Metallicities}
\tablehead{\colhead{Name} & \colhead{Slitmask} & \colhead{SN} & \colhead{v$_{\rm helio}$} & \colhead{$\delta$v$_{\rm helio}$} & \colhead{T$_{\rm eff}$} & \colhead{$\delta $T$_{\rm eff}$} & \colhead{$\log (g)$} & \colhead{[Fe/H]} & \colhead{$\delta $[Fe/H]} & \colhead{[$\alpha $/Fe]} & \colhead{$\delta [\alpha$/Fe]} & \colhead{Member} & & \colhead{RA} & & & \colhead{Decl} & \\  &  & \colhead{$\r{A}^{-1}$} & \colhead{km~s$^{-1}$} & \colhead{km~s$^{-1}$}  & \colhead{K} & \colhead{K} & \colhead{cm~s$^{-2}$}  & \colhead{ } & \colhead{ }  & \colhead{ } & \colhead{ } & \colhead{ } & \colhead{h} & \colhead{min} & \colhead{s} & \colhead{deg} & \colhead{arcmin} & \colhead{arcsec}}
\startdata
PAndAS2422 & and18b & 27.27 & $-$50.16  & 0.84  & 3848.25  & 0.0056  & 0.27    & $-$3.0     & 0.071   & 1.2     & 0.067   & N & 00 & 02 & 14.06 & $+$44 & 55 & 43.8 \\
PAndAS2476 & and18b & 27.59 & 6.93    & 1.19  & 3772.0   & 0.0051  & 0.23    & $-$2.684   & 0.064   & 1.125   & 0.072   & N & 00 & 02 & 01.71 & $+$44 & 55 & 50.1 \\
PAndAS2665 & and18b & 30.46 & $-$110.05 & 22.49 & \nodata  & \nodata & \nodata &  \nodata & \nodata & \nodata & \nodata & N & 00 & 02 & 05.63 & $+$44 & 56 & 06.3 \\
PAndAS3420 & and18b & 23.77 & $-$156.47 & 3.06  & 3644.22  & 0.005   & 0.18    & $-$3.245   & 0.131   & 1.2     & 0.115   & N & 00 & 02 & 12.07 & $+$44 & 57 & 13.9 \\
PAndAS3620 & and18b & 24.12 & $-$11.58  & 1.43  & 3500.0   & 0.0051  & 0.08    & $-$3.256   & 0.112   & 1.2     & 0.093   & N & 00 & 02 & 13.86 & $+$44 & 57 & 30.7 \\
PAndAS4128 & and18b & 4.77  & $-$89.13  & 19.91 & \nodata  & \nodata & \nodata &  \nodata & \nodata & \nodata & \nodata & N & 00 & 02 & 00.52 & $+$44 & 58 & 07.3 \\
PAndAS4556 & and18b & 8.79  & $-$17.69  & 40.98 & \nodata  & \nodata & \nodata &  \nodata & \nodata & \nodata & \nodata & N & 00 & 02 & 11.52 & $+$44 & 58 & 37.8 \\
PAndAS5740 & and18b & 18.71 & $-$35.9   & 2.15  & 3552.6   & 0.0048  & 0.26    & $-$2.436   & 0.127   & 0.53    & 0.167   & N & 00 & 02 & 00.25 & $+$45 & 00 & 06.3 \\
PAndAS5837 & and18b & 14.23 & $-$91.43  & 3.74  & \nodata  & \nodata & \nodata &  \nodata & \nodata & \nodata & \nodata & N & 00 & 02 & 02.48 & $+$45 & 00 & 13.4 \\
PAndAS5861 & and18b & 24.67 & $-$4.82   & 1.47  & 3500.0   & 0.0051  & 0.08    & $-$3.528   & 0.174   & 1.2     & 0.132   & N & 00 & 02 & 13.25 & $+$45 & 00 & 15.5\\
\enddata
\tablecomments{This table is available in its entirety in machine-readable form.}
\end{deluxetable*}
\end{rotatetable*}

%% file: 2_main.tab
\begin{deluxetable*}{@{\extracolsep{12pt}}clllcc}
\label{tab:results}
\tablecolumns{6}
\tablecaption{Adopted and Derived\tablenotemark{a} Parameters of \andxviii}
\tablehead{
\colhead{Property} &\colhead{39 stars\tablenotemark{b} (57)}&
\colhead{38 stars, (56)}& 
\colhead{38 stars, [$\alpha$/H]}&
\colhead{Priors} & \colhead{Units}
}
\startdata
R.A.\ center                 &\multicolumn{3}{c}{0.560417$_{-0.00940}^{+0.00940}$} & & degrees\\
Dec.\ center                 &\multicolumn{3}{c}{45.08889$_{-0.00193}^{+0.00193}$} & & degrees\\
Distance                     &\multicolumn{3}{c}{1.33$_{-0.88}^{+0.88}$}           & & Mpc \\
\hline
\multicolumn{6}{c}{Chemical evolution models} \\
\hline
$\peff$ (Leaky Box)         & $5.4_{-0.8}^{+1.0}$     & $4.9_{-0.8}^{+1.0}$ & $7.6_{-1.2}^{+1.6}$  & $0.0-100.0$ & $10^{-2}$*Z$_{\mathSun}$\\
$\peff$ (Pre-Enriched)      & $5.0_{-0.8}^{+1.0}$     & $4.5_{-0.8}^{+0.9}$ & $5.3_{-1.1}^{+1.4}$  & $1.0-25.0$ & $10^{-2}$*Z$_{\mathSun}$\\
${\rm [Fe/H]}_{0}$ (Pre-Enriched)   & $-3.20_{-1.62}^{+0.55}$       & $-3.03_{-1.53}^{+0.44}$   & $-2.04_{-0.31}^{+0.19}$  & $-5.0--0.5$ & \\ 
$\peff$ (Ram PS)            & $0.13_{-0.08}^{+0.25}$        & $0.24_{-0.12}^{+0.15}$    & $0.17_{-0.09}^{+0.19}$  & $0.0-0.5$ & Z$_{\mathSun}$\\
$z_{\rm str}$ (Ram PS)  & $0.17_{-0.09}^{+0.79}$     & $0.11_{-0.04}^{+0.03}$ & $0.21_{-0.14}^{+0.59}$  & $1.52*10^{-3}-1.52$ & $10^{-2}$ \\
$\zeta_{\rm str}$ (Ram PS)  & $3.4_{-2.0}^{+2.3}$        & $4.6_{-2.0}^{+1.6}$    & $1.1_{-0.7}^{+0.6}$  & $0.0-7.0$ & \\ 
$\peff$ (Accretion)         & $5.4_{-0.7}^{+0.7}$     & $4.7_{-0.5}^{+0.7}$ & $7.2_{-0.9}^{+0.9}$  & $0.0-50.0$ & $10^{-2}$*Z$_{\mathSun}$\\
M$_{\rm Acc}$ (Accretion)   & $3.5_{-1.3}^{+2.1}$        & $3.4_{-1.2}^{+2.2}$    & $6.9_{-2.9}^{+2.1}$  & $1.0-10.0$ & \\ 
\hline
\multicolumn{6}{c}{Kinematic models} \\
\hline
$\langle\vheliotex\rangle$ (no rotation)  & $-337.3_{-1.3}^{+1.4}$      & \multicolumn{2}{c}{$-337.2_{-1.4}^{+1.5}$}      & trial~run & km s$^{-1}$\\
$\sigma_{v}$ (no rotation)                & $9.8_{-0.9}^{+1.1}$         & \multicolumn{2}{c}{$10.0_{-1.0}^{+1.1}$}        & trial~run & km s$^{-1}$\\
$\langle\vheliotex\rangle$ (with rotation)& $-337.3_{-1.4}^{+1.4}$      & \multicolumn{2}{c}{$-337.3_{-1.4}^{+1.4}$}      & trial~run & km s$^{-1}$\\
$\sigma_{v}$ (with rotation)              & $9.9_{-1.0}^{+1.1}$         & \multicolumn{2}{c}{$10.0_{-1.0}^{+1.2}$}        & trial~run & km s$^{-1}$\\
$\vrot$                                   & $1.7_{-1.2}^{+1.7}$        & \multicolumn{2}{c}{$1.8_{-1.2}^{+1.7}$}         & trial~run & km s$^{-1}$\\
$\theta_{\rm axis}$                       & $3.9_{-1.6}^{+1.3}$         & \multicolumn{2}{c}{$3.9_{-1.6}^{+1.3}$}         & trial~run & rad\\
\enddata
\tablenotetext{a}{The center coordinates and the distance were taken from~\citet{mak17}. The metallicity of the Sun was taken to be 0.0152 (see \citealt{gir02}). Values were taken from corner plots obtained after 500 (10000) burn-in steps and a thinning parameter of 50 (100) for {\it emcee} for chemical evolution models (velocity models). Errors cover the $68\%$ confidence interval.}
\tablenotetext{b}{The sample of 56 stars requires membership based on kinematics and the sodium line criterion.  The sample of 39 stars further excludes stars with high metallicity errors ($\delta$[Fe/H]$<0.3$). The next column is for the same members excluding the possible M31 halo star (PAndAS9634). The last column is for chemical evolution models parameters for the [$\alpha$/H] distribution. Prior edges for $\zeta$ and $z_s$ in the RPS model were taken as $0-2$ and $z_{\mathSun}*10^{-5} - z_{\mathSun}$, respectively.}
\end{deluxetable*}

%% file: 3_mcmc.tab
\begin{deluxetable*}{l|c|c|ccc|ccc|ccc}
\label{tab:mcmcwalkers}
\tablecolumns{12}
\tablecaption{MCMC Parameters}
\tablehead{\colhead{Model} & \colhead{N$_{\rm par}$} & \colhead{N$_{\rm walk}$} & \colhead{N$_{\rm it}$} & \colhead{$log(L)$} & \colhead{AICc} & \colhead{N$_{\rm it}$} & \colhead{$log(L)$} & \colhead{AICc} & \colhead{N$_{\rm it}$} & \colhead{$log(L)$} & \colhead{AICc}}
\startdata
\multicolumn{3}{c}{} & \multicolumn{3}{c}{39 (57) members, [Fe/H]} & \multicolumn{3}{c}{38 (56) members, [Fe/H]} & \multicolumn{3}{c}{38 members, [$\alpha$/H]} \\
\hline
Velocity           & 2    & 5     & $10^5$      & -212.42 & 429.06 & $10^5$      & -209.23 & 422.69  &\multicolumn{3}{c}{$-$}\\
Velocity (r.)      & 4    & 8     & $10^6$      & -211.97 & 432.71 & $10^6$      & -208.78 & 426.34  &\multicolumn{3}{c}{$-$}\\
Leaky Box          & 1    & 10    & $12000^{a}$ & -316.97 & 636.04 & 12000       & -306.90 & 615.91  & 11800       & -307.35 & 616.81\\
Pre-Enriched       & 2    & 40    & 12300       & -315.56 & 635.85 & 11600       & -305.19 & 614.73  & 15000       & -301.89 & 608.13\\
Accretion          & 2    & 10    & 4300        & -314.32 & 632.97  & 4500       & -303.80 & 611.95  & 4200        & -301.76 & 607.86\\
RPS                & 3    & 30    & 40900       & -316.71 & 640.11 & 28000       & -300.75 & 608.20  & 29700       & -307.15 & 621.00\\
\enddata
\tablenotetext{a}{Leaky Box and Pre-Enriched models were run for $10^{4}$ more iterations.}
\tablecomments{The columns show (1) the number of unknown parameters in the model, (2) the number of walkers (chains) in the MCMC, (3) the number of iterations or links in the chain, (4) the maximum logarithm of the likelihood, and (5) the corrected Akaike information criterion (Sections~\ref{sec:mcmc},~\ref{sec:reschem}).}
\end{deluxetable*}

%% file: 4_comparison.tab
\begin{deluxetable*}{ccccccc}
\tablecolumns{7}
\tablecaption{Properties of Andromeda XVIII.\label{tab:comparison}}
\tablehead{
\colhead{Research} & \colhead{$\alpha$, $\delta$} & \colhead{$m-M$} & \colhead{R$_{\rm helio}$, kpc} & \colhead{R$_{\rm M31}$, kpc} & \colhead{r$_{\rm 1/2}$, arcmin} & \colhead{$\log (r_{\rm 1/2}/kpc)$} 
}
\startdata
\protect\citet{mcc08} & 00~02~14.5$^{+10}_{-10}$~~45~05~20$^{+10}_{-10}$ & $25.66^{+0.13}_{-0.13}$ & 1355$^{+88}_{-88}$ & 589 & 0.92$^{+0.05}_{-0.06}$ & $-0.44^{+0.036}_{-0.041}$ \\
\protect\citet{tol12} & $-^a$ & $-$ & 1355$^{\rm +83.6, a}_{-78.8}$ & 590.9$^{\rm +87.5, a}_{-82.6}$ & $-$ & $-0.38^{\rm +0.03, a}_{-0.03}$ \\
\protect\citet{con12} & $-^a$ & $25.42^{+0.07}_{-0.08}$ & 1214$^{+40}_{-43}$ & $457^{+39}_{-47}$ & $-$ & $-$\\
\protect\citet{col13}, Section 4 & $-^a$ & $-$ & 1214$^{+40, a}_{-43}$ & $-$ & $-$ & $-0.49^{+0.031, a}_{-0.033}$ \\
\protect\citet{mak17} & $-^a$ & $25.62^{+0.09}_{-0.17}$ & 1330$^{+60}_{-90}$ & 579.$^{+87.}_{-87.}$ & $-$ & $-$ \\
\protect\citet{woj20} & $-^a$ & $-$ & $-$  & $-$ & $-$ \\
This work & $-^a$ & $25.62^{\rm +0.09, a}_{-0.17}$ & 1330$^{\rm +60, a}_{-90}$ & 579$^{\rm +87., a}_{-87.}$ & 0.92$^{\rm +0.05, a}_{-0.06}$ & $-0.45^{+0.032}_{-0.07}$ \\
\enddata
\end{deluxetable*}

\begin{deluxetable*}{ccccccccc}
\tablecolumns{9}
\tablehead{
  \colhead{$[Fe/H]$ (oldest)} & \colhead{$[Fe/H]$ (inter)} & \colhead{$N_{\rm mem}$} 
& \colhead{$v_{\rm helio}$} & \colhead{$\sigma _v$} & \colhead{$\log (M_{\rm 1/2}/M_{\mathSun})$}
& \colhead{E$_{B-V}$ used} & \colhead{M$_{V}$} & \colhead{$\log (L_{\rm 1/2}/L_{\mathSun})$}
}
\startdata
\multicolumn{2}{c}{$-1.8^{+0.1}_{-0.1}$} & $-$  & $-$ & $-$  & $-$ & 0.104 & $<-9.7$ & $-$ \\
\multicolumn{2}{c}{$-$} & 22  & $-332.1^{+2.7}_{-2.7}$ & 9.7$^{+2.3}_{-2.3}$ & $7.43^{+0.24}_{-0.24}$ & $-$ & $-9.7^{\rm +0.1, a}_{-0.1}$  & $5.49^{\rm +0.03, a}_{-0.03}$ \\
\multicolumn{2}{c}{$-$} & $-$  & $-$ & $-$ & $-$ & 0.104 & $-$  & $-$ \\
\multicolumn{2}{c}{$-1.4^{+0.3}_{\rm -0.3}$} & 4 & $-346.8^{+2.0}_{-2.0}$ & 0.0$^{+2.7}$ & 0$^{+6.15}$ & $-$ & $-9.7^{a}$ & $-$ \\
$-1.74^{+0.20}_{-0.20}$ & $-1.33^{+0.20}_{-0.20}$ & $-$ & $-332.1^{\rm +2.7, a}_{-2.7}$ & 9.7$^{\rm +2.3, a}_{-2.3}$ & 6.62$^{\rm +0.03, b}_{-0.03}$ & 0.093 & $-10.41^{+0.28}_{-0.28}$ & $-$ \\
\multicolumn{2}{c}{$-1.33^{+0.02}_{\rm -0.02}$} & 20 & $-332.1^{\rm +2.7, a}_{-2.7}$ & $-$ & 5.9$^{\rm +0.3, b}_{-0.3}$ & $-$ & $-$  & $-$ \\
\multicolumn{2}{c}{$-1.49^{+0.36}_{-0.36}$} & 38 & $-337.2^{+1.5}_{-1.4}$ & 10.0$^{+1.1}_{-1.0}$ & $7.38^{+0.12}_{-0.17}$ & $-$ & $-$  & $-$ \\
\enddata
\tablenotetext{a}{Values adopted from other works.}
\tablenotetext{b}{Total measured stellar mass.}
\end{deluxetable*}